\documentclass[pr,twocolumn,showpacs]{revtex4}
\usepackage{amsmath}
\usepackage{amssymb}
\usepackage{bm}
\usepackage{epsfig}
\usepackage{graphicx}
\usepackage{color}

\def\dfrac{\displaystyle\frac}

\newcommand{\sign}{\mathop{\rm sign}}

\newcount\timehh  \newcount\timemm
\timehh=\time \divide\timehh by 60
\timemm=\time
\begin{document}

\title{Coherent spin dynamics of electrons and holes in semiconductor
quantum wells and quantum dots under periodical optical excitation: resonant spin amplification versus spin mode-locking}

\author{I.~A. Yugova$^{1,2}$, M.~M. Glazov$^{3}$, D.~R. Yakovlev$^{1,3}$, A.~A. Sokolova$^{2}$, and M. Bayer$^{1}$}

\affiliation{$^1$ Experimentelle Physik 2, Technische Universit\"at
  Dortmund, 44221 Dortmund, Germany}

\affiliation{$^2$ Physical
Faculty of St.Petersburg State University, 198504 St. Petersburg,
Russia}

\affiliation{ {$^3$ Ioffe Physical-Technical Institute, Russian
Academy of Sciences, 194021 St. Petersburg, Russia}}

\date{\today, file = \jobname.tex, printing time = \number\timehh\,:\,\ifnum\timemm<10 0\fi \number\timemm}

\begin{abstract}

The coherent spin dynamics of resident carriers, electrons and holes, in semiconductor quantum structures is studied by periodical optical excitation using short laser pulses and in an external magnetic field. The generation and dephasing of spin polarization in an ensemble of carrier spins, for which the relaxation time of individual spins exceeds the
repetition period of the laser pulses, are analyzed theoretically. Spin polarization accumulation is manifested either as resonant spin amplification or as mode-locking of carrier spin coherences. It is shown that both regimes have the same origin, while their appearance is determined by the optical pump power and the spread of spin precession frequencies in the ensemble.

\end{abstract}
\maketitle

\section{Introduction}\label{sec:intro}

The coherent spin dynamics of carriers in semiconductor nanostructures attract considerable attention nowadays due to future quantum information technologies based on spintronics applications~\cite{Awschalom_Spintronics, Gabi_book_1, Fritz_book}. With respect to fundamental studies this research field delivers exciting and unexpected results on the properties of spin systems and the possibility to control them by external fields or by structural parameters.

Optical pump-probe techniques for time-resolved measuring of Faraday and Kerr rotation are based on excitation by trains of laser pulses where the pulse durations range from hundreds of femtoseconds to a few picoseconds. They have been demonstrated to be among the most reliable tools for investigating coherent spin
dynamics~\cite{Awschalom_Spintronics, Dyakonov_Spin, Kikkawa98, Kikkawa_Science97, GaN,kennedy:045307,zhu07,QW,Gupta_PRB99,Gupta_PRB02,Petroff_APL01,
A.Greilich07212006, greilich06,Greilich_PRB07}.
The principle of these magneto-optical techniques is the following: an intense laser pulse of circularly polarized light (the pump) is used to orient spins and, therefore, to create a macroscopic spin polarization~\cite{OptOr}. This polarization is probed by the linearly polarized probe pulses through rotation of their polarization plane after propagation through the spin polarized medium (the Faraday effect) or reflection at this medium (the Kerr effect).  The probe pulse is time-delayed relative to the pump pulse, and by tuning this delay one can measure the spin polarization dynamics. To study the coherent spin dynamics the sample is exposed to an external magnetic field, typically oriented perpendicular to the light wave vector (Voigt geometry), which allows one to detect the precession of the optically induced spin polarization and monitor its decay. Application of these techniques to single spins, which is potentially possible~\cite{Ber08,Atature07}, is demanding. Studying spin ensembles that contain millions of carrier
spins is much more convenient~\cite{A.Greilich07212006, greilich06, Gaby book}.

In pump-probe experiments the spin dynamics evolution is typically measured over times shorter than the repetition period of the pump
pulses, which is about 13~ns for commonly used mode-locked Ti:Sapphire
lasers emitting pulses at a repetition rate of $75-80$~MHz. It has been shown experimentally that in bulk semiconductors, quantum wells (QWs) and
quantum dots (QDs) the carrier spin relaxation time can substantially
exceed the repetition period~\cite {Awschalom_Spintronics,
 Dyakonov_Spin}. In this case the spin polarization induced by
subsequent pump pulses can accumulate if a phase
synchronization condition is fulfilled for the precessing carrier spins. It results in two effects: resonant spin amplification (RSA), observed in bulk and QW spin systems with a relatively small dispersion of precession
frequencies, and spin mode-locking (SML) found for
an ensemble of singly-charged QDs with a large dispersion of Larmor frequencies (see,
e.g., Refs.~[\onlinecite{Gaby book, chapter6}] and references
therein).

For studying the RSA regime experimentally, scanning the magnetic field
has been suggested instead of the commonly used scan of the pump-probe time delay~\cite{Kikkawa98}, used also for tracing SML. The probe pulse arrival time in this case is fixed at a small negative delay prior to the pump pulse. The resulting RSA spectrum is a periodic function of magnetic field from which information such as carrier $g$ factor and dephasing time of the spin ensemble can be extracted.

In this paper we show that the RSA and SML are two different manifestations of the same phenomenon: spin accumulation caused by the periodic excitation with pump pulse trains. We elaborate the
fundamental differences in conditions for appearance of these two
regimes. The most important parameters in this regard are
the pump power and the spin precession frequency spread
causing spin dephasing.  Differences of the two parameters, in turn, lead to different phenomenologies in experiment, providing significantly different capabilities for analyzing spin systems quantitatively.

The paper is organized as follows. In Section \ref{sec:I} we recall the basic concepts and equations for describing spin coherence generation. We discuss the difference between the classical and quantum mechanical approaches to describing carrier spin coherence generation for resonant trion excitation. Then we consider generation of long-lived spin coherence during the trion lifetime. We describe the spin dynamics of charged carriers and trions in magnetic field and discuss the effects of spin relaxation and spin precession of the trion spin on the long-lived spin coherence of resident carriers. We also consider here the long-lived dynamics after generation and the spin accumulation caused by the train of pump pulses.
Section \ref{sec:rsa} is devoted to the RSA regime, for which we consider different conditional effects: trion spin relaxation, nuclear field fluctuations, and spin relaxation anisotropy. The conditions, which are important for observing RSA, and the characteristics, which one can extract from the analysis of RSA signals, are collected at the end of  Sect.~\ref{sec:rsa}. Section
\ref{sec:modelock} describes the main features of mode-locking of electron spin coherences. Then in Section \ref{sec:rsa_vs_ml} we compare the spin dynamics in the RSA and SML regimes,  obtain conditions for the SML regime and discuss the transition to the RSA regime.
In the Conclusions, we give a comparative description of the RSA and
SML regimes and their applicability to investigations of
long-lived spin dynamics in low-dimensional systems.

\section{Generation of spin coherence}\label{sec:I}

In the following we analyze the long-lived spin coherence of resident
carriers (electrons and holes) generated by periodic light excitation
in semiconductor quantum wells and quantum dots. We consider a
situation with a low concentration of resident carriers, when
the probability to have two charge carriers with significantly overlapping wavefunctions is low. In this case, mainly
few-particle complexes, excitons (electron-hole pairs) and trions
(three particle complexes) can be optically excited, while other
many-body correlations are negligible.
For quantum wells this corresponds to typical carrier densities smaller
than 10$^{10}$~cm$^{-2}$, for which at liquid Helium temperatures
carriers are localized on QW width fluctuations with respect to their
in-plane motion. Only one carrier per localized site is typical for
such concentrations and the distance between the localized carriers
exceeds the extensions of neutral and charged exciton wavefunctions. For quantum dots the low concentration regime corresponds to occupation of a dot with only one resident carrier, i.e. to a regime of singly-charged QDs.

Here we consider the theoretical aspects of the problem. We
do not discuss the experimental aspects of the observations (measurements) of long-lived spin coherences and features of ellipticity and Faraday rotation signals.  We limit ourselves to the degenerate
pump-probe regime, when the probe laser has the same photon energy as the pump one, and to resonant excitation of the trion states. We assume that the pulse duration is significantly shorter than all characteristic relaxation times of the considered spin system. Other regimes are studied in detail elsewhere\cite{yugova09,fokina-2010,longpulse11,reviewFTT}.
These conditions are typical for experiments with semiconductor nanostructures~\cite{Awschalom_Spintronics,  Dyakonov_Spin}.

For low concentrations of resident carriers charged excitons
(trions) play an important role in the generation process of carrier
spin coherence~\cite{greilich06,zhu07}. A negatively charged exciton
(T$^-$ trion) is a bound state of two electrons and one
hole, while a positively charged exciton (T$^+$ trion) is a bound state of two holes and one electron. The trion ground state at zero magnetic field has a singlet spin configuration, such that the spins of the two
identical carriers are aligned opposite to each other and the trion
Zeeman splitting is controlled by the $g$ factor of the unpaired
carrier, e.g., the hole in T$^-$. Hereinafter we assume
that only heavy holes with angular momentum projections
$\pm 3/2$ onto the growth axis are involved.

The theoretical analysis used in this paper can be equally applied to structures with resident electrons or resident holes. In order
to do that we have introduced universal notations: the resident carrier
spin $\bm S$, the trion spin $\bm S^T$, the Larmor frequency of the resident carrier $\omega = g\mu_B B/\hbar$, and the Larmor frequency of the trion $\Omega = g_T\mu_B B/\hbar$. Here $\bm B$ is the external magnetic field, $\mu_B$ is the Bohr magneton, $g$ and $g_T$ are the $g$ factors of the resident carrier and trion, respectively. Similarly
  universal notations are also used in what follows to denote
  characteristic time scales.

In $n$-type doped structures with resident electrons, $\bm S$ is the
electron spin, $\bm S^T$ is a (pseudo) spin of the T$^-$ trion
($S^T=+1/2$ for $+3/2$ hole and $-1/2$ for $-3/2$ hole),
$\omega$ is the electron Larmor frequency, and $\Omega$ is the T$^-$
Larmor frequency determined by the hole $g$ factor.
Correspondingly, in $p$-type doped structures with resident
holes, $\bm S$ is the heavy hole pseudospin, $\bm S^T$ is the spin of the T$^+$ trion, which corresponds to the electron spin in this
trion, $\omega$ is the heavy hole Larmor frequency, and
$\Omega$ is the T$^+$ Larmor frequency determined by the electron $g$ factor.

For the sake of simplicity we consider in most parts of this paper
$n$-type doped structures with resident electrons, as there are more experimental data available for these structures. Wherever we analyze $p$-type doped structures this will be notified. Before we proceed to the analysis of spin precession in magnetic field and spin dephasing processes, let us inspect briefly the models of optical generation of spin coherence.

\subsection{Resonant excitation of trion. Classical and quantum
  mechanical approaches to carrier spin coherence
  generation}\label{subsec:classqnt}

The important quantity for spin coherence generation in the system is the singlet trion resonance that is excited optically. The trion generation probability for resonant excitation depends on the light polarization and the spin orientation of the resident carrier. For instance, in $n$-type doped structures a $\sigma^+$ polarized pump generates a hole with
spin projection $+3/2$ onto the light propagation axis $z$ and an
electron with spin projection $-1/2$. Therefore, trion
formation is possible only when the resident electron has
spin projection  $+1/2$. As a result, the circularly polarized pump
pulse selects electrons with particular spin orientation from the
ensemble of resident electrons to form trions. This, in turn, leads to
spin polarization of the resident electrons.

\begin{figure}[hptb]
\includegraphics[width=1.\linewidth]{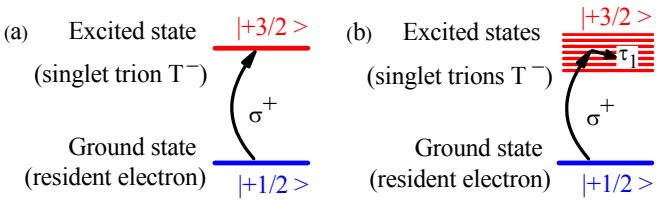}
\caption{ (a) Scheme of transitions for a strongly localized electron
  (e.g., in a singly-charged quantum dot). The initial state for the
  optical transition is a resident electron and final state is a
  singlet trion T$^-$. This scheme is consistent with the quantum
  mechanical approach.  {(b)} Scheme of transitions for the case of
  weakly localized resident carriers (e.g., in a quantum well with a low
  density electron gas); $\tau_1$ denotes the scattering
  time between different trion states. This scheme is consistent
  with the classical approach.}
\label{fig:qnt_class}
\end{figure}

There are two approaches for describing the spin
coherence generation by circularly polarized light pulses~\cite{reviewFTT}. The first one is essentially quantum mechanical: a singly-charged QD or a QW with a localized resident electron is modeled as a two-level
system~\cite{shabaev:201305,greilich06,kennedy:045307,yugova09}. The
ground state corresponds to the resident electron, while the excited
state is the singlet trion, see Fig.~\ref{fig:qnt_class}(a).

The interaction of the two-level system with the resonant pump pulse
depends on the pulse parameters (polarization, intensity and pulse
duration) and on the level occupations.
The pump pulse action time, $\tau_p$, is assumed
to be the shortest of all timescales in the problem, namely the trion
dephasing and scattering times, the electron Larmor precession period,
the trion radiative lifetime, the spin
dephasing/decoherence times, etc. Under usual experimental conditions
the trion lifetime is much shorter than the pump pulse repetition
period and, consequently, trion spin polarization is absent shortly before the next pump pulse, i.e. is not detectable at negative time delays.
It follows then, that the resident carrier spin pseudovector $\bm
S=(S_x, S_y,S_z)$ before the pump pulse, $\bm S^b$, and after the pump
pulse, $\bm S^a$ are related to each other through~\cite{yugova09}:
\begin{subequations} \label{pm}
\begin{eqnarray}
S_z^a &=&   \pm \frac{Q^2-1}{4} +\frac{Q^2+1}{2}S_z^b\:, \label{szpm}\\
S_x^a &=& Q\cos{\Phi} S_x^b \pm Q\sin{\Phi} S_y^b\:,\label{sxpm} \\
S_y^a &=& Q\cos{\Phi} S_y^b \mp Q\sin{\Phi} S_x^b\:,\label{sypm}
\end{eqnarray}
\end{subequations}
where the signs $\pm$ correspond to $\sigma^+$ and $\sigma^-$ polarized
pump pulses in $n$-type structures and to $\sigma^-$ and $\sigma^+$ pulses for $p$-type structures, respectively. This sign definition is also valid for Eqs.~\eqref{Jz}, \eqref{small:power} and \eqref{sz:max}.  The parameters
$0\leqslant  Q \leqslant 1$ and $0\leqslant \Phi < 2\pi$ characterize the
pump pulse area and the spectral detuning of the pulse from the trion
resonance. The explicit expressions for these quantities
are given in Ref.~[\onlinecite{yugova09}]. For the resonant pump pulse
$\Phi=0$ and $Q=\cos{(\Theta/2)}$, where $\Theta$ is the pump pulse
area: $\Theta = \int 2|\langle d
\rangle E(t)|dt / \hbar$. Here $\langle d \rangle$ is the dipole
transition matrix element and $E(t)$ is the smooth envelope
of the electric field of the
laser pulse. The $z$ component of the trion spin
pseudovector after, e.g., a $\sigma^+$ pump pulse in
a $n$-type system or a $\sigma^-$ pump pulse in a $p$-type system is
given by~\cite{yugova09}
\begin{equation}
\label{Jz}
S^T_z = S_z^b - S_z^a = \frac{1-Q^2}{4}\left(2S_z^b \pm 1\right).
\end{equation}

Such an approach has been proven to be appropriate for the description of spin coherence generation in $n$-type singly-charged QDs~\cite{greilich06}. At low pump powers, where $\Theta \ll 1$, the additive contribution to the
electron spin $z$ component equals to
\begin{equation}
\label{small:power}
S_z^a-S_z^b=\mp \frac{\Theta^2}{{16}} \propto P,
\end{equation}
where $P$ is the pump pulse power.
One of the main predictions of the considered quantum mechanical
approach is that for high pump powers the electron spin $z$ component
depends periodically on the pump area, i.e., shows Rabi oscillations
inherent to a two-level system, see
e.g. Refs~[\onlinecite{bonadeo98,A.Greilich07212006,greilich06}].

The experimentally studied situation in $n$-type QW
structures is different~\cite{zhu07}. For low pump powers and resonant
trion excitation the electron spin coherence increases linearly with
the pump power, see Eq.~\eqref{small:power}, while at high powers the
spin $z$ component saturates and Rabi oscillations are not
observed~\cite{zhu07}. Clearly, the two-level model is not sufficient
for describing such a behavior. The most probable reason is related to the weaker localization of electrons and trions in quantum wells and,
hence, to the presence of many trion states. Scattering
between these states becomes possible, as schematically illustrated in
Fig.~\ref{fig:qnt_class}(b). The optical coherence of the trion with the
pump is lost due to this scattering, while spin coherence is
preserved. As a result, if the scattering time between different trion
states, $\tau_1$, is considerably shorter than $\tau_p$, the Rabi
oscillations at high pump powers vanish~\cite{zhukov10}, because the
population of the excited state of the two-level system coupled to the
optical pulse is small. At the same time, the spin polarization
generated by the pump pulse can be substantial, because spin does not
relax during scattering. With an increase of the pump pulse
power the electron spin saturates at the value
\begin{equation}
\label{sz:max}
S_{z,{\rm max}} = \mp N/4,
\end{equation}
where $N$ is the total number of resident electrons in the system. The
amount of trions formed for resonant excitation of the initially
unpolarized electron ensemble cannot exceed $N/2$, since
only half of the resident electrons have suitable spin orientation
to become excited to trion singlets. The other $N/2$ of the resident electrons, which are not captured to trions, have become fully polarized.

As will be shown below in Sec.~\ref{subsec:rsa_fast}, the quantum
mechanical and classical approaches give the same results at low pump
powers. Subsequently, we will use the quantum mechanical
approach because it gives good descriptions for spin
coherence generation for QDs in any excitation power regime and for QWs in the low power excitation regime.

{\subsection{Generation of long-lived spin coherence during the trion
    life time. Spin dynamics of charged carriers in magnetic field}

\subsubsection{Spin dynamics of resident carrier and
  trion}\label{sec:electrontrion}

 Right after the excitation pulse the coupled dynamics of resident
 carrier spin, $\bm S$, and trion spin, $\bm S^T=(S^T_x,S^T_y,S^T_z)$,
 can be described by the following system of
 equations~\cite{shabaev:201305,greilich06,zhu07,yugova09}:
\begin{subequations}
\label{system}
\begin{equation}
\frac{d\bm S^T}{dt} = \frac{\mu_B}{\hbar} [g_T \bm B \times \bm S^T] - \frac{\bm S^T}{\tau_s^T} - \frac{\bm S^T}{\tau_r}, \label{J}
\end{equation}
\begin{equation}
\frac{d\bm S}{dt} = \frac{\mu_B}{\hbar} [g \bm B \times \bm S] - \frac{\bm S}{\tau_s} + \frac{S^T_z\bm e_z}{\tau_r}. \label{S}
\end{equation}
\end{subequations}
Here $\bm e_z$ is the unit vector along the $z$ axis. The magnetic
field $\bm B$ is assumed to be parallel to the $x$ axis. $\tau_s^T$ is the trion spin relaxation time, $\tau_s$ is the phenomenological spin
relaxation time of the resident carrier \cite{comment}, and $\tau_r$
is the trion radiative lifetime. It is worth to mention that
carriers left behind after trion recombination are polarized parallel
or antiparallel to the $z$ axis due to the optical selection rules, see the last term $\propto S^T_z\bm e_z$ in Eq.~\eqref{S}.

From Eqs.~\eqref{system} the carrier spin projection onto the magnetic field, $S_x$, is conserved. Introducing the trion spin
lifetime, $\tau_T = \tau_s^T\tau_r/(\tau_s^T+\tau_r)$,  we arrive at
the following expression for the transverse carrier spin component
$S_+ = S_z + \mathrm i S_y$~\cite{greilich06}:
\begin{multline}
\label{S+}
S_+(t) = S_{+,0} \mathrm e^{-\mathrm i \omega t - t/\tau_s}
  \\  + S^T_{z,0}\left[-\xi \mathrm e^{-\mathrm i \omega t - t/\tau_s}
    + \mathrm e^{-t/\tau_T} (\xi \cos{\Omega t}  + \chi \sin{\Omega t}
    )\right].
\end{multline}
Here the subscript $0$ denotes the spin components at time $t=0$, when the pump pulse is finished, e.g., $S_{+,0} = S_z(0)+\mathrm i S_y(0)$.
\begin{eqnarray}
\label{xichi}
\xi &=& \xi_1+ \mathrm i \xi_2 = \frac{\mathrm i \omega/\gamma-1}{\gamma\tau_r[(1- \mathrm i \omega/\gamma)^2+(\Omega/\gamma)^2]}, \\
\quad \chi &=& \chi_1+ \mathrm i \chi_2 = \frac{\Omega/\gamma}{\gamma\tau_r[(1- \mathrm i \omega/\gamma)^2+(\Omega/\gamma)^2]},
\end{eqnarray}
and $\gamma = \tau_T^{-1} - \tau_s^{-1}>0$.

In order to have a closed equation system~\eqref{system}, we have to relate the carrier and trion spins at $t=0$. This can be done through
Eqs.~\eqref{pm} and \eqref{Jz}. After a single pump pulse ($\bm
S^b=0$) one has
\[
S^T_{z,0} = - S_{z,0}.
\]
The first term in the right hand side of
Eq.~\eqref{S+} describes the carrier spin precession. The term
proportional to $S^T_{z,0}\mathrm e^{-\mathrm i \omega t}$ describes the spin polarization of the
resident carrier after trion recombination. Below, we
consider the relation of these two contributions as a function of spin
system parameters and external conditions.

\subsubsection{Effect of trion spin relaxation on spin coherence of
  resident carrier}\label{subsec:holetrion}

In absence of an external magnetic field the efficiency of
resident carrier spin coherence generation is solely determined by the
trion spin
relaxation~\cite{kennedy:045307,PhysRevB.75.115330,zhu07}. This becomes clear from Eq.~\eqref{S+}, which for $B=0$ reduces to
\begin{equation}
\label{Sz:nofield}
S_z(t) = S_{z,0} \mathrm e^{ - t/\tau_s} + S^T_{z,0} \xi \left(-\mathrm e^{ - t/\tau_s} + \mathrm e^{-t/\tau_T}\right).
\end{equation}
It follows from Eq.~\eqref{xichi} that $\xi=-(\tau_r \gamma)^{-1}
\approx -(1+\tau_r/\tau_s^T)^{-1}$, provided that the carrier spin
relaxation time exceeds by far both trion recombination time and trion
spin lifetime. These conditions are readily fulfilled in
experiment. Hence, the long-lived carrier spin coherence is given by
\begin{equation}
\label{Sz:nofield1}
S_z(t) = (S_{z,0}-S^T_{z,0}\xi)\mathrm e^{- t/\tau_s}, \quad t\gg  \tau_T.
\end{equation}
If spin relaxation in the trion is suppressed, i.e. $\tau_s^T\gg
\tau_r$, then $\xi \to -1$. Therefore, since for a single pump pulse
$S^T_{z,0} = -S_{z,0}$, the contribution of the carrier left behind
from the trion decay compensates exactly the spin polarization of the
remaining, non-excited carrier component. As a result, no long-lived spin coherence for resident carriers is generated. In general, when the resident carrier has been polarized before pump pulse arrival, this carrier polarization will not be affected by the pump pulse and conserved
after trion recombination.  To conclude, trion spin relaxation is required to give rise to a non-zero long-lived spin coherence of the resident carriers in absence of a magnetic field.

\subsubsection{Spin precession of resident carrier}

The carrier spin precession about an external magnetic field results in an imbalance of resident and returning spins. Hence, long-lived spin
coherence can be excited even in the absence of trion spin relaxation.
Provided that the trion spin does not precess~\cite{com5},
$\Omega=0$, the long-lived carrier spin coherence is given
by~\cite{zhu07}
\begin{equation}
\label{Sz:field}
S_z(t) = \sign(S_{z,0})|S_{z,0}-S^T_{z,0}\xi|\mathrm e^{- t/\tau_s} \cos{(\omega t -\varphi)}, \quad t\gg  \tau_T
\end{equation}
where $\varphi$ is the initial phase, which can be related to the
parameter $\xi$, see Ref.~[\onlinecite{zhu07}] for details. Note, that
in Ref.~[\onlinecite{zhu07}] the phase is shifted by $\pi/2$ with respect
to our definition in Eq.~\eqref{Sz:field}. The amplitude of the
long-lived spin coherence $\mathcal A_s$ after a single pump pulse can
be recast as
\begin{equation}
\label{ASz:field}
\mathcal A_s=|S_{z,0}-S^T_{z,0}\xi| = |S_{z,0}(1+\xi)| \approx |S_{z,0}| \frac{|\omega\tau_r|}{{\sqrt{1+(\omega\tau_r)^2}}},
\end{equation}
where the latter approximate equality is valid for a long trion spin
relaxation time fulfilling the relation $\tau_s^T\gg \tau_r$. According to Eq.~\eqref{ASz:field} the long-lived spin amplitude first
increases with growing magnetic field $\propto \omega \tau_r$ and then saturates in strong fields.

The general case of arbitrary $\omega\tau_r$ and $\tau_r/\tau_s^T$
is illustrated in Fig.~\ref{fig:fig2}.
Panel (a) demonstrates the dependence of the
long-lived spin coherence amplitude $\mathcal A_s$ on magnetic field
(expressed as $\omega/\gamma$) for different
values of the ratio $\tau_r/\tau_s^T$. 
Depending on the parameter $\tau_r/\tau_s^T$, the change
of amplitude $\mathcal A_s$ as function of magnetic field (through
$\omega\sim B$) occurs for different field values since $\gamma$
itself is determined by $\tau_r$ and $\tau_s^T$.

\begin{figure}[hbt]
\includegraphics[width=0.9\linewidth]{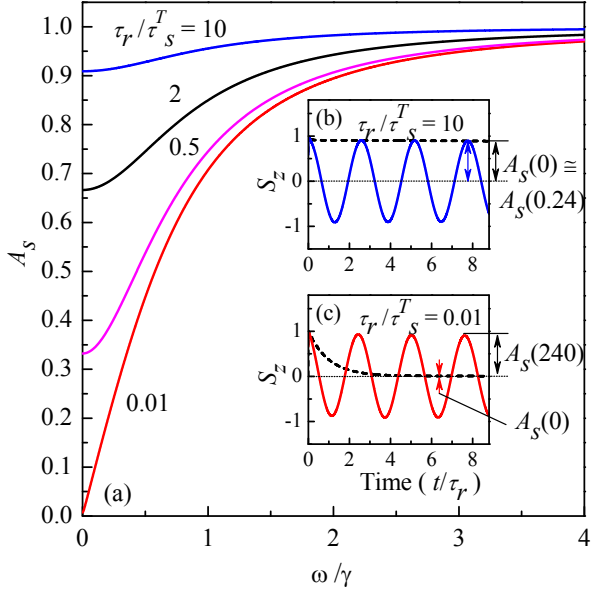}
\caption{(Color online) (a) Dependence of the long-lived spin
  coherence amplitude $\mathcal A_s$ on the carrier Larmor precession
  frequency for different values of $\tau_r/\tau_s^T$. (b),(c) Carrier
  spin coherence $S_z(t)$ normalized to $S_z(0)$ for two different
  values of $\tau_r/\tau_s^T$. The spin dynamics at zero magnetic field are shown by the dashed lines. The solid lines show $S_z(t)$ at finite magnetic field ($\omega\tau_r=2.4$). The arrows show the corresponding amplitudes
  $\mathcal A_s(\omega/\gamma)$ for these conditions.} \label{fig:fig2}
\end{figure}

The panels (b) and (c) in Fig.~\ref{fig:fig2} show the carrier
spin coherence $S_z(t)$ calculated for fast ($\tau_r/\tau_s^T=10$) and
slow ($\tau_r/\tau_s^T=0.01$) spin relaxation of the trion. The solid  and dashed
lines show $S_z(t)$ in zero and finite magnetic field,
respectively. One can see from Fig.~\ref{fig:fig2}(b), that at
$\tau_r/\tau_s^T=10$ the amplitude of the long-lived spin
coherence ($t \gg \tau_r$) in magnetic field coincides with the one at
$B$ = 0. In the graph this corresponds to the coincidence of the dashed line (zero field) with the maxima of the oscillating
solid line (finite field, $\omega/\gamma=0.24$). In other words, the
application of magnetic field here does not change the efficiency of
spin coherence generation. This is, however, not the case for the
smaller ratio of $\tau_r/\tau_s^T=0.01$
($\omega/\gamma=240$). As one can see in
Fig.~\ref{fig:fig2}(c) the dashed line at longer delays has
considerably smaller amplitude than the maxima of the solid line,
$\mathcal A_s(0) \ll \mathcal A_s(240)$. This means that the
amplitude of long-lived spin coherence, $\mathcal A_s$, can be
strongly increased by external magnetic fields. To conclude, even in
the absence of spin relaxation in the trion the application of an external magnetic field leads to appearance of long-lived spin polarization of the resident carriers.

\subsubsection{Effect of spin precession in trion on spin coherence of
  resident carrier}\label{subsec:holetrion:prec}

Spin precession of the trion, characterized by the frequency $\Omega$,
also provides a mechanism for generating long-lived
carrier spin coherence.
Although the in-plane hole $g$ factor in quantum wells and in self-assembled quantum
dots is rather small~\cite{marie99,yugova02,stevenson}, the spin precession of the hole in the T$^-$ trion may become important in tilted magnetic fields~\cite{Machnikowski10}, and in the case of the T$^+$ trion excited in $p$-doped structures~\cite{korn_njp}.

Allowing for $\Omega \ne 0$ results in the following expression for
the amplitude of the long-lived spin coherence $\mathcal A_s$
[c.f. Eqs.~\eqref{Sz:field} and \eqref{ASz:field}]:
\begin{equation}
\label{ASz:field:Omega}
\mathcal A_s=|S_{z,0}-S^T_{z,0}\xi| = |S_{z,0}(1+\xi)| \approx |S_{z,0}| \frac{(\Omega\tau_r)^2}{1+(\Omega\tau_r)^2},
\end{equation}
where in the latter equality we assume a trion spin relaxation time, $\tau_s^T\gg \tau_r$, and neglect the resident carrier spin
precession, $\omega\ll \Omega$. It follows from
Eq.~\eqref{ASz:field:Omega} that the spin precession in the trion acts
similar to the trion spin relaxation.
Here it does not matter whether the spin of the unpaired carrier in
the trion was rotated by the magnetic field or flipped due to spin
relaxation: in both cases long-lived carrier spin polarization arises.

\begin{figure}[hbt]
\includegraphics[width=0.9\linewidth]{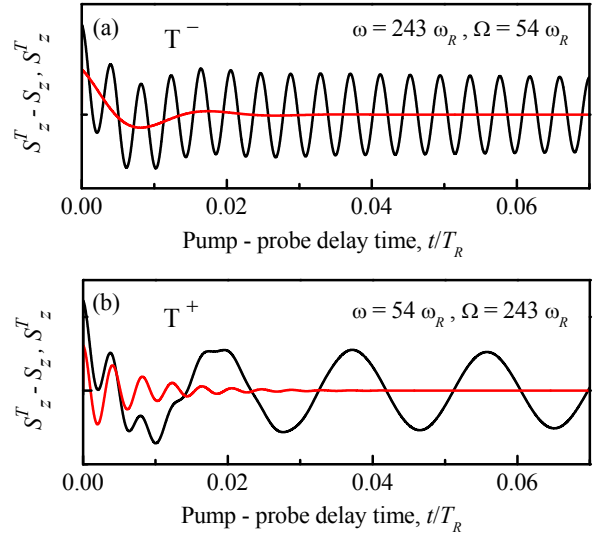}
\caption{(Color online) Spin dynamics of resident carriers and trions
  for (a) negatively charged trions T$^-$, $n$-type and (b)
  positively charged trions T$^+$, $p$-type. The black curves show the
  temporal evolution of $S^T_z - S_z$. The gray (red) curves give only
  the trion contribution to the signal, $S^T_z$.} \label{fig:fig3}
\end{figure}

The situation becomes richer when spin precession of both resident carrier and trion occurs.  Figure~\ref{fig:fig3} shows the spin dynamics of T$^-$ trion and resident electron [panel (a), $n$-type] and T$^+$ trion and resident hole [panel (b), $p$-type]. The black curves give the difference $S^T_z-S_z$ which corresponds to signals commonly measured in experiment, see also Eqs.~(54) and (57) in
Ref.\cite{yugova09}. It is clearly seen that at short delay times not exceeding the trion lifetime the $S^T_z-S_z$ dynamics are additionally modulated by the trion Larmor frequency. For clarity, the spin dynamics of trions, $S^T_z$, are shown separately by the gray (red) lines. They decay relatively fast being limited by the trion recombination. The trion radiative lifetimes as well as spin
relaxation times are taken the same in both panels:
$T_R/\tau_r=130$, $T_R/\tau_s^T =13$ and $T_R/\tau_s =
2.6$. The carrier and trion Larmor precession frequencies are given in the panels. $T_R$ is the
repetition period of excitation pulses and $\omega_R=2\pi/T_R$. For
commonly used mode-locked lasers with a repetition frequency of
75~MHz $T_R=13.3$~ns.

\subsection{Spin accumulation induced by a train of pump pulses}
\label{subsec:accum}

In experiments on coherent spin dynamics periodic trains of pump
pulses are commonly used. When the spin relaxation time of the
resident carrier is comparable or longer than the repetition period of
the pump pulses, i.e. $\tau_s > T_R$, the
steady-state carrier spin polarization results from the
cumulative contribution of multiple pump pulses. In external magnetic
fields applied in the Voigt geometry,
the steady-state situation is reached for each precessing
spin by relatively long trains of
pump pulses: the decay of the spin polarization is then balanced by the
pumping. As a result, the carrier spin after each repetition period,
$\bm S(T_R)$, given by Eq.~\eqref{S+}, should be equal to the carrier
spin right before the pump pulse arrival, which we denote by $\bm S^b$ (see
Fig.~\ref{fig:fig4}). Using the connection between the carrier spins
before and after the pump pulse, Eq.~\eqref{pm}, and assuming that the
pump pulse is resonant with the trion transition, $\Phi=0$, one
immediately comes to the following expression for the carrier spin $z$
component before pump pulse arrival:
 \begin{widetext}
\begin{equation}
\label{S-z}
S_{z}^b(\omega) = \pm \frac{1}{2}\frac{K}{1+Q\mathrm e^{-2T_R/\tau_s} - \mathrm e^{-T_R/\tau_s}(1+Q)\cos(\omega T_R) - K},
\end{equation}
where the signs $\pm$ correspond to different polarizations
of optical pumping and different types of resident carriers, cf. Eqs.~(\ref{pm}),
and
\[
K=\frac{(1-Q^2)\mathrm e^{-T_R/\tau_s}}{2} \left\{ (1+\xi_1)[Q\mathrm
  e^{-T_R/\tau_s}- \cos(\omega T_R)] - \xi_2 \sin(\omega T_R)
\right\}.
\]
\end{widetext}

Equation~\eqref{S-z} shows that the spin $z$ component before
the next pump pulse arrival, $S_z^b$, is a periodic function of
magnetic field (see  Fig.~\ref{fig:fig5}) with maxima of $|S_z^b|$ at
frequencies $\omega$ satisfying the phase synchronization condition
(PSC)~\cite{Kikkawa98,beschoten,A.Greilich07212006}:
\begin{equation}
\label{PSC}
\omega=N\omega_R= \dfrac{2\pi N}{T_R }, \quad N=0,1,2, \ldots .
\end{equation}
Here $\omega_R=2\pi/T_R$ is the repetition frequency of the pump pulses.
Indeed, as one can see from
time-resolved signals shown in Fig.~\ref{fig:fig4}, if the spin
precession period of the resident carrier is commensurable with the
pump pulse repetition period, then the spin coherence
generated by the pump is always in phase with that from the
previous pulse [see signal around zero time delay, Fig.~\ref{fig:fig4}(a)], and carrier spin polarization is accumulated. Let this phase, $\phi$, be zero.
Otherwise, if the spin precession and pump repetition periods
are not commensurable, the accumulation of spin polarization is not
efficient, as seen from the comparison of the amplitudes in Figs.~\ref{fig:fig4}(a) and \ref{fig:fig4}(b).

In general, the electron spin precession has a
particular phase, see Eq.~(\ref{Sz:field}), which
we determine here as the difference $(\omega T_R - 2\pi N)$, where $N$ is the
largest integer satisfying the condition $(\omega T_R - 2\pi N) \ge 0$.  The phase can be expressed as:
\begin{eqnarray}
\label{phase}
\cos(\phi)=-S_z^b/\sqrt{(S_z^b)^2+(S_y^b)^2},\\ \nonumber
\sin(\phi)=S_y^b/\sqrt{(S_z^b)^2+(S_y^b)^2}.
\end{eqnarray}

Note, that in Fig.~\ref{fig:fig4} and further on in this paper we show for convenience the inverted signal $-S_z$  (in order to have positive signals for $\sigma^+$ pumping). This sign change does not
affect the obtained results but is more suitable for
their graphic presentation. For an ensemble of resident carriers with different spin precession frequencies, $\omega$, Eq.~\eqref{S-z} should be averaged over their distribution~\cite{glazov08a}, see below Sec.~\ref{subsec:dephasing}
and Eq.~(\ref{Gauss}).

\begin{figure}[hbt]
\includegraphics[width=1.0\linewidth]{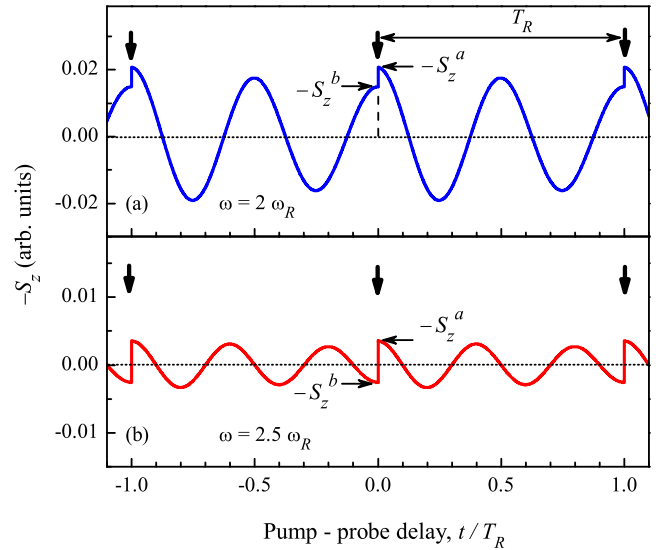}
\caption{(Color online) Dependencies of resident carrier spin polarization
$S_z$ on pump-probe delay for a carrier spin precession frequency which
is (a) commensurable with the pump repetition frequency
$\omega=2\omega_R$ and (b) not commensurable with this frequency
$\omega=2.5\omega_R$. Parameters of calculations are: $\tau_s=3T_R$,
$\Theta=0.1 \pi$. Thick vertical arrows show the arrival times of the pump pulses. Phase $\phi$ of the oscillating polarization, $-S_z$, is
$\phi=0$ in panel (a) and $\phi=\pi$ in panel~(b).} \label{fig:fig4}
\end{figure}

\begin{figure}[hbt]
\includegraphics[width=0.9\linewidth]{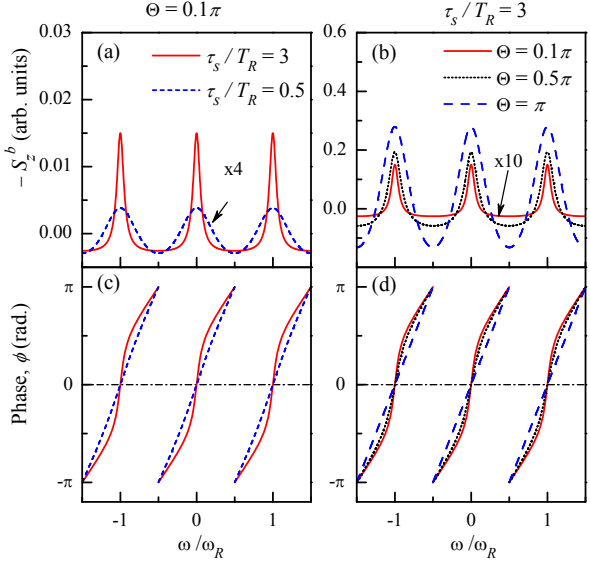}
\caption{(Color online) {Dependence of the resident carrier spin
polarization $S_z^b$ and its phase on magnetic field expressed by
$\omega/\omega_R=g \mu_B B/(\hbar \omega_R)$. Data are shown for
zero time delay (right before the pump pulse arrival), calculated
for different ratios $\tau_s/T_R$ at $\Theta=0.1 \pi$ (a,c) and for
different pump pulse areas $\Theta$ at $\tau_s/T_R = 3$
(b,d).}} \label{fig:fig5}
\end{figure}

Figure~\ref{fig:fig5} shows $S_z^b$
calculated after Eq.~\eqref{S-z} for different pump pulse
areas $\Theta$ in the case of fast trion spin relaxation.
As $x$ axis scale in Fig.~\ref{fig:fig5} we take the
ratio of spin precession frequency $\omega$ and $\omega_R$, which
represents the magnetic field dependence of $S_z^b$ as $\omega
\propto B$. Integer numbers on the $x$ axis correspond to magnetic
fields, for which the spin precession frequency satisfies the PSC of
Eq.~\eqref{PSC}. At these magnetic fields the amplitude of the
resident carrier spin polarization, $-S^b_z$, increases resonantly evidencing favorable conditions for spin accumulation, see
Fig.~\ref{fig:fig5}(a). It is obvious that the accumulation efficiency
is controlled by the factor $\tau_s / T_R$, as the accumulation occurs only when the spin relaxation time of the resident carrier, $\tau_s$,
exceeds considerably the repetition period of the pump pulses. This is
confirmed by the calculations shown in Fig.~\ref{fig:fig5}(a). For
a fixed value of  $\tau_s / T_R$ an increase of the pump pulse area
results in a broadening of the peaks, see Fig.~\ref{fig:fig5}(b).

The phases of the signals from Figs.~\ref{fig:fig5}(a) and
\ref{fig:fig5}(b) are shown in panels (c) and (d),
respectively. One clearly sees that the zeros of the phase correspond
to maxima of spin polarization, $-S_z^b$, and the values $\phi = \pm \pi$
correspond to its minima.

One should note, that the magnitude of the accumulated spin
polarization, as well as the width of the resonant peaks in the
magnetic field dependence of  $-S_z^b$, are determined not only by the
pump pulse power and the carrier spin relaxation time, but also by the
mechanism of long-lived spin coherence generation and the spin
dephasing time. We present the analysis of these effects in the
following Sections.

\section{Resonant spin amplification}
\label{sec:rsa}

We begin with the classical expression for carrier
spin polarization under RSA
conditions~\cite{Kikkawa98,beschoten}. The underlying assumptions are the following:
(i) only carrier spin polarization is considered, and (ii) it is
supposed that each pump pulse generates only a $z$ component
of spin polarization, whose magnitude is $S_0$. All
non-additive effects of the pump pulse~\cite{zhukov10} are disregarded.
After single pump pulse excitation the carrier spin dynamics are
described by a decaying cosine function periodic with the
Larmor precession frequency $\omega$ and decay with time $\tau_s$. The effect of a long train of pump
pulses on the carrier spin polarization can be calculated as:
\begin{equation}
\label{weak:class_rsa_sum}
S_z(\omega, t) = \sum_{k=0}^{\infty}S_0\mathrm e^{-(t
  +kT_R)/\tau_s}\cos[\omega (t +kT_R)],
\end{equation}
where $t$ is the pump-probe delay and $k=0,1,2,...$ .
This equations can be rewritten~\cite{beschoten,glazov08a} as:
\begin{multline}
\label{weak:class_rsa}
S_z(\omega, t) = \frac{S_0}{2}  e^{-t/\tau_s} \times \\
\frac{ e^{-T_R/\tau_s}\cos(\omega t)-\cos\left[\omega (t +T_R)\right]}{\cosh(T_R/\tau_s)-\cos(\omega T_R)}.
\end{multline}
It follows from Eq.~(\ref{weak:class_rsa}) that for sufficiently long
decay times $\tau_s \gtrsim T_R$ the carrier spin
has sharp resonances as function of
magnetic field. As will be shown by our calculations,
this corresponds to the solid line in Fig.~\ref{fig:fig5}(a) and gives the RSA signals presented in Fig.~\ref{fig:fig6}. The peak positions at zero pump-probe delay correspond to spin precession frequencies which are
commensurable with the pump repetition frequency
$\omega_R=2\pi/T_R$. The expression~(\ref{weak:class_rsa}) near
commensurable frequency ($|\omega T_R - 2\pi N| \ll 1$) and at a zero
time delay can be written as:
\begin{equation}
\label{Lorentian0}
S_z^{b} \sim \frac{1}{(\omega T_R - 2\pi N)^2 + (T_R/\tau_s)^2},
\end{equation}
Here we assume that $T_R/\tau_s \ll 1$. The peak width is determined by the relaxation time of the electron
spin polarization. Note, that for the spin ensemble the time $\tau_s$
should be changed to the dephasing time $T_2^*$ \cite{comment_t2}. This
allows one to measure spin relaxation
and spin dephasing times exceeding $T_R$,
i.e., for conditions where direct determination by time-resolved methods becomes inapplicable.
The equations~(\ref{weak:class_rsa}) and~(\ref{Lorentian0})
describe a number of experiments well, see, e.g.,
Refs.\cite{Kikkawa98,beschoten,zhu06,ast08}, and facilitate evaluation of carrier $g$ factors and spin dephasing times~\cite{comment_t2}.

However, one sees that the spin polarization in
Eqs.~(\ref{weak:class_rsa}) and~(\ref{Lorentian0}) increases
to infinity if $\tau_s$ becomes larger
  and larger. Moreover, such an approach disregards completely the
  spin dynamics of trions and the specifics of carrier spin
  dephasing in external magnetic fields. This case requires
  a special treatment. There are also experiments which reveal a
  complicated shape of RSA spectra or a complete absence of RSA
  despite of very long spin relaxation times, which cannot be
described by this simple model~\cite{korn_njp,sokolova09,greilich09}. The general analysis required for such cases is presented below.

\subsection{Fast spin relaxation in trion}
\label{subsec:rsa_fast}

If the spin relaxation of the unpaired carrier in the trion is fast, $\tau_s^T\ll \tau_r$, the trion spin dynamics does not affect the spin polarization of the resident carrier, see Sec.~\ref{subsec:holetrion}. 
In this case the carrier polarization induced by the pump pulse is not compensated by the carriers left after trion recombination, as these carriers are unpolarized.
Then $\xi=0$ and the parameter $K$ in Eq.~\eqref{S-z} has the simple form~\cite{A.Greilich07212006,yugova09}
\begin{equation}
\label{K}
K=\frac{(1-Q^2)\mathrm e^{-T_R/\tau_s}}{2} \left[ Q\mathrm e^{-T_R/\tau_s}-\cos(\omega T_R)  \right].
\end{equation}
The detailed analysis of Eqs.~\eqref{S-z} and \eqref{K} for this case
is given in Refs.~[\onlinecite{A.Greilich07212006,yugova09}]. If,
moreover, the pump pulse area $\Theta$ is small, so that $1-Q \ll 1$,
Eq.~\eqref{S-z} together with Eq.~\eqref{K} go over into the classical
expression of Eq.~\eqref{weak:class_rsa} for carrier spin polarization under RSA conditions.

It follows, that for frequencies near the phase synchronization
condition of Eq.~\eqref{PSC}, the spin $z$ component of the resident
carrier can be recast as \cite{yugova09}:
\begin{equation}
\label{Lorentian}
S_z^{b} \sim \frac{1}{(\omega T_R - 2\pi N)^2 + \left[T_R/\tau_s + (1-Q)\right]^2},
\end{equation}
where we assume that $T_R/\tau_s\ll 1$, $1-Q\ll 1$ and $|\omega T_R - 2\pi N| \ll 1$. One sees from Eq.~\eqref{Lorentian} that the RSA peak width is determined by $T_R/\tau_s$ or $1-Q$, whichever is larger.

Figure~\ref{fig:fig6} shows RSA signals calculated for a small pump power, $\Theta=0.1 \pi$, at two different delays. The shape of the RSA signal at large negative delay ($t=-0.1T_R$) differs from the one at zero delay due to the different phases of the spin precession.

An increase of the pump pulse area results in broadening of the RSA
peaks, as was already shown in Fig.~\ref{fig:fig5}(b). For increasing pump pulse area the RSA peaks are no longer Lorentzians and, therefore, $S_z^b$ cannot be described by Eq.~\eqref{Lorentian}. The spin polarization for $\Theta=\pi$ and $\tau_s/T_R = 3$ shown in Fig.~\ref{fig:fig5}(b) looks similar to the one for $\Theta=0.1 \pi$ and $\tau_s/T_R = 0.5$ in Fig.~\ref{fig:fig5}(a). Hence, under
strong excitation the dependence of carrier spin polarization on magnetic field becomes cosine-like due to saturation effects. In this case it is not possible to extract the carrier spin relaxation or the dephasing times from the width of RSA peaks.

\begin{figure}[hbt]
\includegraphics[width=1.0\linewidth]{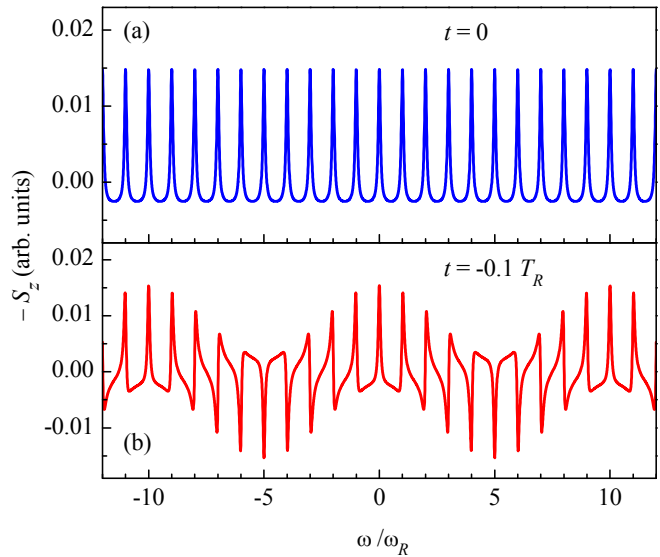}
\caption{(Color online) Carrier spin polarization $S_z$ as function of magnetic field ($\omega \propto B$) at two different pump-probe delays denoted in each panel. $t=0$ means that the signal is calculated for very small negative delay, just before the pump pulse arrival. Parameters of calculations: $\tau_s=3T_R$, $\Theta=0.1 \pi$.} \label{fig:fig6}
\end{figure}

\subsection{Slow spin relaxation in trion: effect of trion spin dynamics}
\label{subsec:rsa_slow}

Let us now turn to the general case, in which the trion spin relaxation
time can be comparable or even longer than its recombination time. It
is instructive to start from the situation in which $\tau_s^T\gg \tau_r$
and long-lived spin coherence appears only due to carrier or
trion spin precession about the magnetic field. Clearly, the peaks in
the $S_z^b(\omega)$ dependence are suppressed for $\omega\tau_r,
\Omega\tau_r \ll 1$ due to inefficient spin generation, and
they increase significantly with an
increase of magnetic field. This is illustrated in
Fig.~\ref{fig:fig7}, where the calculated RSA
signals are shown for $\tau_s^T=30 \tau_r$. Note, that
such unusual RSA spectra with suppression of the peak amplitudes in
weak magnetic fields have been observed experimentally in both
\textit{n}-type and \textit{p}-type
QWs~\cite{sokolova09,fokina-2010,korn_njp,Kugler11}.

Figure~\ref{fig:fig7}(a) shows the signal calculated in
    absence of trion spin precession ($\Omega=0$) shortly
  before the pump pulse arrival. The peak amplitude at zero magnetic field ($\omega=0$) is given by the ratio $\tau_r$ and $\tau_s^T$ and
  goes to zero for infinite $\tau_s^T$. The increase of peak
  amplitudes with increasing magnetic field depends on $\xi$ and,
  therefore, on the ratio $\omega / \gamma$, similar to the amplitude
  dependencies in Fig.~\ref{fig:fig2}. The peak shapes at zero delay differ from being Lorentzian, see {for comparison} Eq.~\eqref{Lorentian}
  and Figs.~\ref{fig:fig5}(a,b) and \ref{fig:fig6}(a), because the spin
  left behind after trion recombination changes the phase of the carrier
  spin precession~\cite{zhu07}.

\begin{figure}[hbt]
\includegraphics[width=1.0\linewidth]{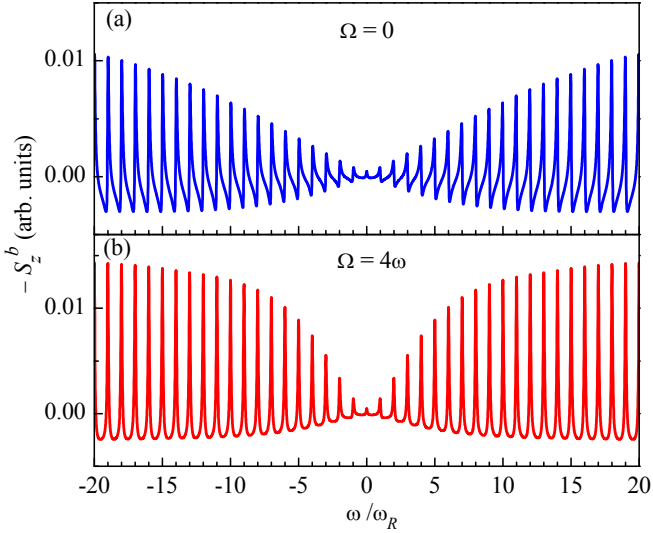}
\caption{(Color online) Impact of slow spin relaxation of the unpaired
  spin in the trion: RSA signals at zero delay without ($\Omega=0$) and
  with ($\Omega=4\omega$) trion spin precession (panels (a) and (b),
  respectively).  Parameters of calculations: $\tau_s^T=30 \tau_r$,
  $\tau_r=0.01 T_R$, $\tau_s=3T_R$ and $\omega \tau_r=4.4$ at $B=1$~T
  and $\Theta=0.1\pi$.
}
\label{fig:fig7}
\end{figure}

It is worth to stress, that we can use the same system of
equations \eqref{system} to describe the spin dynamics in $n$-type
(resident electron and T$^-$ trion) and $p$-type (resident hole and
T$^+$ trion) structures.
Figure~\ref{fig:fig7}(a) illustrates the situation that is typical for
\textit{n}-type QWs~\cite{com5,sokolova09}, in which trion spin
precession is absent. Figure~\ref{fig:fig7}(b) shows the RSA signal with
a trion spin precession frequency $\Omega=4\omega$, which may correspond
to the T$^+$ trion case in \textit{p}-type QWs~\cite{korn_njp,Kugler11}. The
analysis shows that small $\Omega$, i.e.
$\Omega \le \omega$, leads to no significant changes
of the RSA signal shape as compared with one in Fig.~\ref{fig:fig7}(b). A fast precession of the trion spin results in
a faster appearance of long-lived spin coherence with increasing
magnetic field, compare Figs.~\ref{fig:fig7}(a) and \ref{fig:fig7}(b). 

\subsection{Effect of spin relaxation anisotropy}
\label{subsec:rsa_any}

To make our analysis of RSA complete, we
briefly discuss here another effect, which is relevant for weak
magnetic fields. It addresses the situation in which the carrier spin relaxation or the dephasing times are anisotropic. Spin relaxation anisotropy is an inherent feature of semiconductor quantum
wells~\cite{dyakonov86,averkiev:15582,averkiev06,larionov:033302,
  willander}. For simplicity,
we consider the case, in which the $z$ and $y$ spin components of
the resident carriers relax at different time constants, $\tau_{s,z}$
and $\tau_{s,y}$, respectively. Provided that
the long-lived carrier spin coherence is excited by the train of
weak pump pulses, the dependence of the carrier spin $z$ component on the
precession frequency is given by~\cite{glazov08a}
\begin{equation}
\label{S-z:aniso}
S_z^b(\omega) \sim  \frac{\mathcal C(\tilde \omega T_R) - \mathrm
  e^{-T_R/\tau_s}}{\cosh{(T_R/\tau_s)} - \cos{(\tilde\omega T_R)}},
\end{equation}
where
\begin{equation}
\label{taus:eff}
\frac{1}{\tau_s}=\frac{1}{2}
\left(\frac{1}{\tau_{s,z}}+\frac{1}{\tau_{s,y}}\right), \quad \tilde
\omega = \sqrt{\omega^2 - \frac{1}{4}\left(\frac{1}{\tau_{s,z}} -
    \frac{1}{\tau_{s,y}}\right)^2},
\end{equation}
and
\[
\mathcal C(\tilde \omega T_R) = \cos{(\tilde \omega T_R)} -
\frac{1}{2\tilde{\omega}}\left(\frac{1}{\tau_{s,z}} -
  \frac{1}{\tau_{s,y}}\right) \sin{(\tilde \omega T_R)}.
\]

\begin{figure}[hptb]
\includegraphics[width=0.49\textwidth]{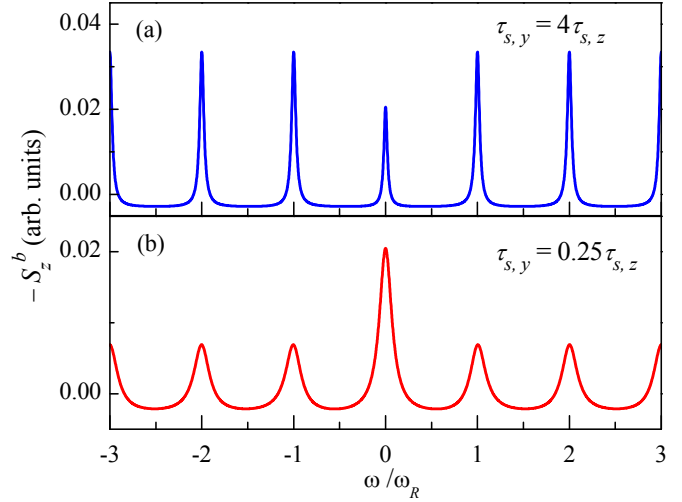}
\caption{Effect of an anisotropy of the carrier spin relaxation times. The electron spin $z$
  component right before pump pulse arrival $(t=0)$ is calculated as function of magnetic field after
  Eq.~\eqref{S-z:aniso}. $\tau_{s,z}=4T_R$. }
\label{fig:rsa:aniso}
\end{figure}

The dependence of carrier spin polarization, $-S_z^b$, on magnetic
field is shown in Fig.~\ref{fig:rsa:aniso} for two cases of
anisotropic carrier spin relaxation: (a) $\tau_{s,y}=4\tau_{s,z}$ and (b)
$\tau_{s,y}=0.25\tau_{s,z}$. The amplitudes of all maxima except the
one at zero field are the same, because they are determined by the
effective spin relaxation time, $\tau_s$, defined by
Eq.~\eqref{taus:eff}. The amplitude of the zero-field peak is
different from the other peaks. If $\tau_{s,y}>\tau_{s,z}$, it is
smaller as compared with the others. The carrier spin relaxation in
absence of a magnetic field is governed solely by $\tau_{s,z}$ and is
faster than at finite magnetic fields, so that accumulation of carrier
spin polarization is weaker at $B=0$. In the opposite case of
$\tau_{s,y}<\tau_{s,z}$ the zero-field peak is higher, because the
lifetime of the spin $z$ component is longer in absence of
magnetic field so that spin accumulation is more
efficient~\cite{Griesbeck11}.

\subsection{Spin decoherence and dephasing}
\label{subsec:dephasing}

The spin relaxation time of localized carriers can be extremely long
reaching up to microseconds for electrons in QDs, for example
~\cite{Ignatiev05}. This is related with quenching of the orbital
motion and the corresponding suppression of spin relaxation mechanisms
contributed by spin-orbit coupling~\cite{PhysRevB.64.125316,PhysRevB.66.161318}.
The coherence time of an individual spin is typically much
longer compared with the spin dephasing time of an inhomogeneous
spin ensemble. The inhomogeneity, which leads to a spread of carrier
spin precession frequencies, results in spin dephasing
characterized by the $T^*_{2}$ dephasing time.
This time measured, e.g., from the decay of spin beats in external
magnetic field is in the few nanoseconds range for QD
ensembles~\cite{A.Greilich07212006,greilich06,carter:167403} and in
the tens of nanoseconds range for QWs containing diluted carrier
gases~\cite{zhu07,ast08,PhysRevB.75.115330,fokina-2010}.

One of the main origins for the inhomogeneity of a spin
ensemble is related to the $g$ factor spread of localized
carriers. For electrons the $g$ factor variation can arise from changes of the effective band gap for different localization
sites~\cite{ivchenko05a,PhysRevB.75.245302,A.Greilich07212006}. For
localized holes the variations are mainly related to changes in the mixing of heavy-hole and light-hole states \cite{Kotlyar01}.
The spread of $g$ factors in a spin ensemble, $\Delta g$, is translated into a spread of spin precession frequencies, $\Delta \omega_g$, and, therefore, results in a spin dephasing rate~\cite{ast08,glazov08a}
\begin{equation}
\label{t2:dg}
\frac{1}{T_{2,\Delta g}^*} \sim \frac{\Delta g \mu_B B}{\hbar}\equiv
\Delta \omega_g,
\end{equation}
which is accelerated with increasing magnetic field.

Another origin of spin dephasing typical for electrons
is related to random nuclear fields in the quantum dots~\cite{merkulov02}. Each localized electron is subject to a hyperfine field of a particular nuclear spin
fluctuation, $\bm B_{\rm n}$, and, therefore, precesses about this field at a frequency $\omega_{\rm n}$. These fluctuations are different for
localization sites, causing dephasing of the electron spin
ensemble. The dephasing rate can be estimated by the root mean square of
the electron spin precession frequency in the field of frozen
nuclear fluctuations~\cite{merkulov02}:
\begin{equation}
\label{t2:dn}
\frac{1}{T_{2,\rm n}^*} \sim \sqrt{\langle \omega^2_{\rm n} \rangle}.
\end{equation}
Assuming a normal distribution of $\bm B_{\rm n}$
Eq.~\eqref{t2:dn} can be rewritten as:
\begin{equation}
\label{t2:dn2}
\frac{1}{T_{2,\rm n}^*} \sim \frac{g \mu_B \Delta_B}{\hbar}\equiv
\Delta \omega_{\rm n}.
\end{equation}
where $\Delta_B$ is the dispersion of the nuclear spin
fluctuation distribution \cite{merkulov02}.

Estimates show that $T_{2,\rm n}^*$ is on the order of several
nanoseconds for GaAs quantum dots \cite{merkulov02,syperek11}. Hence,
in weak magnetic fields (e.g., $B\lesssim 0.3$~T for $g = 0.5$ and
$\Delta g = 0.005$~\cite{greilich09}) the spin beat decay for
resident electrons is determined by the hyperfine interaction, and in
higher fields the dephasing is caused by the spread of $g$ factors~\cite{comment_dephasing}.

In quantum wells with a diluted electron gas the electron localization
on well width fluctuations is considerably weaker compared to the
QD case. As a result, $\Delta g$ is smaller and the hyperfine
interaction is weaker. Therefore, the spin dephasing times can reach
$\sim 30 \div 50$~ns in weak magnetic fields and at low
temperatures~\cite{zhu07,fokina-2010}. Below the effect of a spin precession frequency spread on RSA
signals is analyzed.

\subsubsection{Spread of $g$ factors}
\label{subsec:spread}

For a more realistic approach we need to take into account the precession
frequency spread, $\Delta \omega$, in the spin ensemble. Here for
distinctness we consider only the frequency spread caused by $\Delta g$ (the spread related with the nuclear spin fluctuations is considered below). For ensemble of carrier spins with a spread of $g$ factors, $\Delta g$, the spread of Larmor precession frequencies, $\Delta \omega_g$, is proportional to the magnetic field:
\begin{equation}
\label{deltag}
\Delta \omega_g(B)=\Delta g \mu_B B/\hbar
\end{equation}
To model the ensemble RSA signal one has to sum the contributions of
the individual spins~\cite{glazov08a} over the $g$ factor
distribution function:
\begin{equation}
\label{Gauss}
\rho(g)=\frac{1}{\sqrt{2 \pi} \Delta g} \exp \left[\frac{- (g - g_0)^2}{2(\Delta g)^2} \right],
\end{equation}
where $g_0$ is the average $g$ factor value in the spin
ensemble, resulting in an average Larmor frequency: $\omega_0=g_0 \mu_B B/\hbar$.

RSA spectra calculated by means of Eqs.~\eqref{S-z} and \eqref{Gauss} for short trion spin relaxation, i.e. dependencies of the
carrier spin polarization, $-S_z$, on magnetic field in terms of
$\omega_0 / \omega_R$ are shown in Fig.~\ref{fig:fig9} for two negative time delays. An increase of magnetic field leads to 
broadening of the RSA resonances and decrease of their amplitudes. This reflects the acceleration of the spin dephasing rate $1/T_2^* \sim B$, in accordance with Eq.~\eqref{t2:dg}.

\begin{figure}[hbt]
\includegraphics[width=1.0\linewidth]{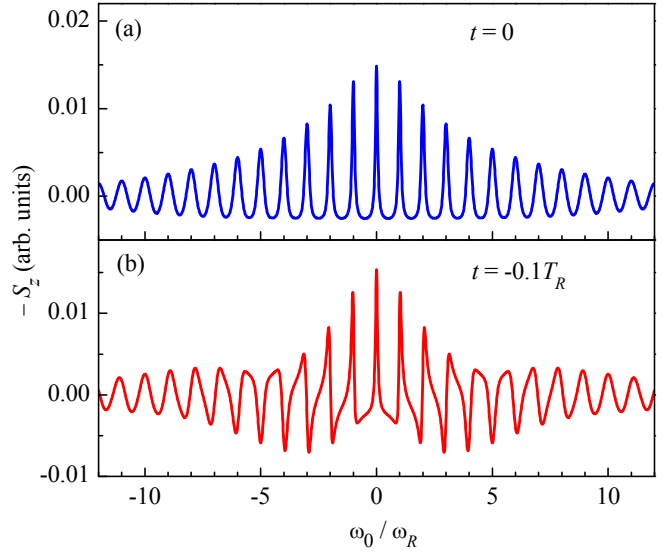}
\caption{(Color online) Dependencies of carrier spin polarization
$S_z$ on magnetic field at two different pump-probe delays $t$
given in each panel and short trion spin relaxation time $\tau_s^T \ll \tau_r$. A frequency spread $\Delta \omega_g = 0.02
\omega_0$, corresponding to $2\%$ dispersion of the carrier $g$ factor, is assumed in the calculations. The dependence on magnetic field is given by $\omega_0/\omega_R=g_0\mu_BB/(\hbar \omega_R)$.} \label{fig:fig9}
\end{figure}

Figures~\ref{fig:fig10}(a) and \ref{fig:fig10}(b) show RSA signals
for long trion spin relaxation, $\tau_s^T=30 \tau_r$, with and
without spin precession in the trion. An ensemble spread of $\Delta
\omega_g = 0.02 \omega_0$
results in a broadening of the RSA peaks and a decrease of their
amplitudes with increasing magnetic field,
similar to Fig.~\ref{fig:fig9}. This results in the
characteristic bat-like shape of the RSA
signal~\cite{sokolova09,fokina-2010,korn_njp,Kugler11} compare with Fig.~\ref{fig:fig7} where the spin dephasing was absent, $\Delta \omega_g=0$. Accounting
for the spread of $\Omega$ does not change the signals significantly.

\begin{figure}[hbt]
\includegraphics[width=1.0\linewidth]{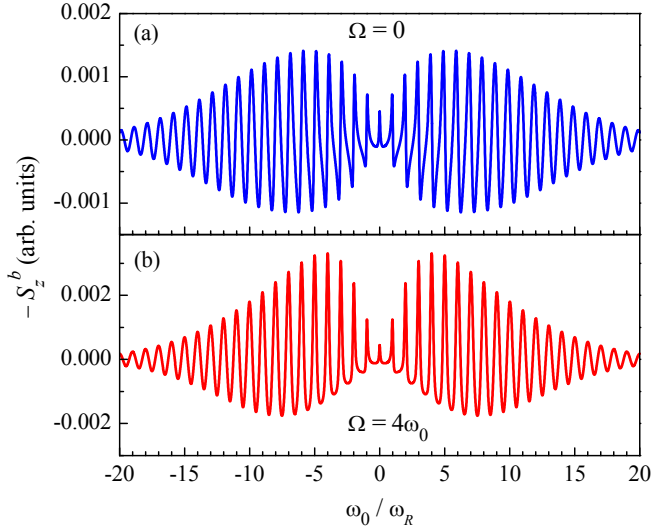}
\caption{(Color online) Effect of slow trion spin relaxation. RSA
  signals at zero delay without [(a) $\Omega=0$] and with [(b)
  $\Omega=4\omega_0$] trion spin precession are shown. The signals are
  calculated assuming a spin precession frequency spread
 $\Delta \omega = 0.02 \omega_0$ of the resident
  carrier. The parameters in the calculations are:
  $\tau_s^T=30 \tau_r$, $\tau_r=0.01 T_R$, $\tau_s=3T_R$ and $\omega_0
  \tau_r=4.4$ at $B=1$~T and $\Theta=0.1\pi$.}
\label{fig:fig10}
\end{figure}

Figure~\ref{fig:fig10}(a) corresponds to a situation that
is obtained for resident electrons oriented by excitation of the T$^-$ trion in $n$-type (In,Ga)As/GaAs QWs~\cite{sokolova09,fokina-2010}. In such structures the in-plane hole $g$ factor is small compared with the
electron $g$ factor, and consequently $\Omega \ll
\omega$, so that the spin precession of the T$^-$ trion can be neglected.

Figure~\ref{fig:fig10}(b) corresponds to
the long-lived hole spin orientation for excitation of the T$^+$ trion in
$p$-type GaAs/(Al,Ga)As QWs~\cite{korn_njp}. For the T$^+$ trion the ratio $\Omega$ and $\omega$ is
opposite, i.e. $\Omega \gg \omega$. In Ref.~[\onlinecite{korn_njp}]
$\Omega=4.5\omega$ and the spin precession in trion affects the RSA signal. 

The results of the calculations
shown in Figs.~\ref{fig:fig10}(a) and \ref{fig:fig10}(b) are in
good agreement with
available experimental data for quantum well structures
\cite{sokolova09,fokina-2010,korn_njp}. All calculations were done for a small pulse area $\Theta=0.1\pi$. The analysis of the case
  of high pump power, which results in saturation effects, shows
  that an increase of the pump power results in an increase of the
signal amplitude and broadening
of all peaks, similar to the case discussed in
Sec.~\ref{subsec:rsa_fast}, see also
Fig.~\ref{fig:fig5}. The bat-like shape of the RSA signal envelope
is conserved even for $\Theta=\pi$ pump pulses.

\subsubsection{Nuclear field fluctuations and resonant spin amplification in weak magnetic fields}
\label{subsec:spread_nuclei}

Interaction of the nuclear spins with hole spins is weak and in
many cases can be neglected. At the same time, for localized
electrons the hyperfine interaction with the nuclei can considerably
contribute to the spin dynamics. Therefore, in this subsection we will
focus on n-type structures containing resident electrons.

In weak magnetic fields the electron spin dephasing time related to the
spread of $g$ factor values, Eq.~\eqref{t2:dg}, proportional to
$1/B$, becomes very long and nuclear field fluctuations play an
important role. The hyperfine fields acting on the electrons due to
these nuclear fluctuations can be as large as $B_{\rm n} \sim 0.5$~mT for
GaAs QWs~\cite{PhysRevB.75.115330} and an order of magnitude larger in
(In,Ga)As QDs~\cite{petrov08}.

For $B\gtrsim B_{\rm n}$ the only important component of the nuclear field fluctuation is the one parallel to the external field $\bm B$. It
results in a spread of Larmor precession frequencies, damping of the
spin beats and broadening of the RSA peaks, provided $B_{\rm n}>
|\Delta g/g| B$.

\begin{figure}[hbt]
\includegraphics[width=0.8\linewidth]{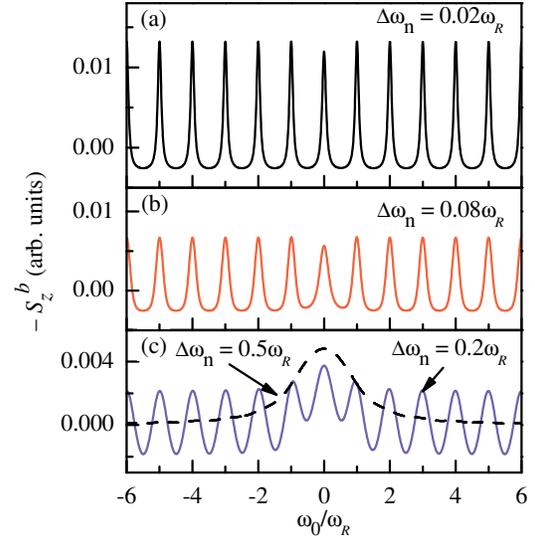}
\caption{(Color online) RSA signals at zero pump-probe delay
  calculated for different spreads of the nuclei fluctuation field
  $\Delta_B$. The frequency spreads ($\Delta \omega_{\rm n} \sim
  \Delta_B$) are given in each panel.  $\Theta=0.1\pi$, $\Delta g =0$
  and $\tau_s^T \ll \tau_r$.} \label{fig:fig11}
\end{figure}

The situation becomes different in weak magnetic fields $B< B_{\rm
  n}$. In this case all components of the nuclear fluctuation field
become important. For illustration we consider a homogeneous electron
spin ensemble  ($\Delta g=0$) in a magnetic field which is the sum of
the external magnetic field $\bm B$ and the fluctuation field $\bm B_{\rm
  n}$. For simplicity, we consider the regime of fast spin
relaxation in the trion ($\tau_s^T \ll \tau_r$). To model the dynamics of the electron spin ensemble one can assume a normal distribution of $\bm B_{\rm n}$:
\begin{equation}
\label{Gauss_nucl}
\rho_n({\bm B_{\rm n}})=\frac{1}{(\sqrt{2 \pi} \Delta_B)^3}
\exp\left(- \frac{{\bm B_{\rm n}}^2}{2(\Delta_B)^2} \right).
\end{equation}
where $\Delta_B$ is the isotropic dispersion of the nuclear fluctuation field distribution ($\Delta_{B,x} = \Delta_{B,y} = \Delta_{B,z}$).
The spread of the Larmor precession frequencies, $\Delta \omega_{\rm
  n}$, does not depend on the external magnetic field:
\begin{equation}
\label{deltan}
\Delta \omega_{\rm n}=g \mu_B \Delta_B/\hbar.
\end{equation}
The average Larmor frequency of the spin ensemble in this case is
equal to the spin precession frequency in an external magnetic field without nuclear fluctuations: $\omega_0=g \mu_B B/\hbar$.

Figure~\ref{fig:fig11} shows RSA
 signals at zero time delay averaged over $B_{\rm n}$ for different
 $\Delta_B$ values. One sees that indeed, an increase of the
frequency spread $\Delta \omega_{\rm n}$ leads to an increase of the
dephasing rate evidenced via broadening of the
RSA peaks. For weak magnetic fields $B < \Delta_B$ the $y$ component of
the nuclear fluctuation field, $B_{\rm n,y}$, which is
perpendicular to $S_z$ and to the external field, can
additionally destroy the long-lived carrier spin polarization. This is
manifested in an additional broadening and a decrease of the
amplitude of the zeroth RSA peak (compared to the $\pm 1$ peaks), as is clearly seen in Fig.~\ref{fig:fig11}(a,b). The enhancement of $S_z$ in the vicinity of zero field for large fluctuations, see
Fig.~\ref{fig:fig11}(c), is due to the fact that the $z$ component of
the spin polarization can not destroy
by a parallel component of the nuclear fluctuation field $B_{\rm n,z}$.

\subsection{Analysis of RSA signals and evaluation of spin
    dephasing times and $g$ factors}

To conclude our analysis of RSA we emphasize that in spite of the possibly complex shape of RSA signals, especially in case of a long spin relaxation in the trion, the analysis allows one to obtain various parameters with high accuracy. This is due to the fact that these parameters are responsible for different features in the RSA spectrum:

\begin{itemize}
\item{the $g$ factor of the resident carriers gives the magnetic field positions of the RSA peaks.}
\item{The $g$ factor spread, {$\Delta g$}, determines the amplitude decrease of the RSA peaks with increasing magnetic field.}
\item{The spin relaxation/dephasing time $\tau_s$ is related to the RSA peak widths~\cite{comment_hanle}}.
\item{The ratio of spin relaxation time $\tau_s^T$ and radiative lifetime $\tau_r$ of the trion determines the possible increase of RSA peak amplitudes with increasing magnetic field. If $\tau_r$ is obtained from an independent time-resolved measurement, then $\tau_s^T$ can be extracted from fitting the RSA spectrum.}
\item{For long spin relaxation in the trion (when the RSA signal has a bat-like shape) the symmetry of the RSA peaks at zero pump-probe delay can indicate the fact that the trion $g$ factor is larger than that of the resident carrier ($|g_T| \gg |g|$). However, the value of the trion $g$ factor should be obtained from another experiment.}
\item{Finally, the amplitude and the width of the zero-field RSA peak {can contain information on the} anisotropy of the spin relaxation of delocalized carriers and the nuclear effects for localized carriers.}
\end{itemize}

The spin dynamics parameters considered above can be extracted only
for sufficiently homogeneous ensembles and at weak excitation powers
(small pump pulse areas), which is typical for semiconductor QWs.

It is worth to mention, that there are other generation
  mechanisms of long-lived spin coherence for nonresonant optical
  excitation~\cite{fokina-2010,zhu07,Kugler11}.
In this case, the RSA signal can change its shape
dramatically. However, a detailed analysis allows
one to identify the generation and
relaxation mechanisms of carrier spin polarization and obtain the
corresponding quantitative information about relaxation processes.

\section{Mode-locking of carrier spin coherences}
\label{sec:modelock}

Now we turn to strongly inhomogeneous spin systems, for which the spread of the spin precession frequencies is so large that
\begin{equation}
\label{strong:inh}
T_2^* <T_R.
\end{equation}
Still, the spin relaxation time of the resident carrier is assumed to
exceed by far the repetition period, $\tau_s \gg T_R$. In this case
the ensemble spin polarization generated by a pump pulse
decays within the time $T_2^*$, i.e., disappears before the next pump
pulse arrival. Figure~\ref{fig:fig12} presents model calculations, which show the dynamics of the carrier spin polarization excited by a train of the pump pulses. Indeed, the polarization decays quite rapidly after the pump pulses, but thereafter reemerges at negative delays $-T_2^*
\lesssim t <0$. Such a behavior has been explained in terms of
mode-locking of carrier spin coherences that are synchronized by the periodic train of pump pulses~\cite{A.Greilich07212006,Gaby book}.

\begin{figure}[hbt]
\includegraphics[width=1.\linewidth]{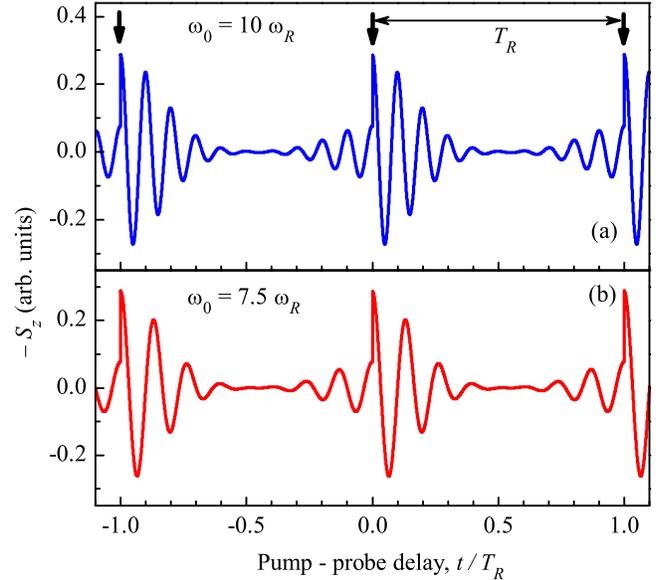}
\caption{(Color online) Carrier spin polarization as function of pump-probe delay for precession frequencies which (a) satisfy the PSC of Eq.~\eqref{PSC} and (b) do not satisfy it.  The frequency spread is $\Delta \omega =\omega_R$ and $\Theta=\pi$. Thick vertical arrows indicate the arrival times of the pump pulses.} \label{fig:fig12}
\end{figure}

If the condition~\eqref{strong:inh} is fulfilled, the pump pulse excites a broad distribution of spin precession frequencies, among which there are several frequencies satisfying the phase synchronization condition of Eq.~\eqref{PSC}. The carrier spins with such precession frequencies are excited much more efficiently, i.e., accumulate more spin polarization than the other ones. As a result, the main contribution to the signal is given by the commensurable spin beat frequencies. In other words, the spins satisfying the PSC become resonantly amplified, while others are not, and the synchronized spins contribute mostly to the experimentally measured signal of carrier spin polarization. Such behavior of the spin signals, characteristic for the mode-locking of carrier spin coherences, has been observed in \textit{n}-type singly-charged (In,Ga)As QDs~\cite{A.Greilich07212006,chapter6,Gaby book}.

The calculations shown in Fig.~\ref{fig:fig12} are carried out after Eqs.~(\ref{S-z}) and (\ref{deltan}) assuming, for simplicity, that the trion spin relaxation is fast, $\tau_s^T \ll \tau_r$, and the spread of the carrier spin precession frequencies $\Delta\omega=\omega_R$ does not depend on the magnetic field strength.

Let us have a closer look on the signals in Fig.~\ref{fig:fig12}. It is remarkable, that the phase of the spin beats before the next pump pulse arrival is fixed for any magnetic field. The average precession frequency of spin ensemble, $\omega_0$, satisfies the PSC in Fig.~\ref{fig:fig12}(a) while it does not in Fig.~\ref{fig:fig12}(b). The phase, however, in both cases is exactly the same and it also coincides with the one after the pump pulse, $\phi=0$. This is in strong contrast with the regime of weak dephasing ($T_2^* \gtrsim T_R$), see Fig.~\ref{fig:fig5}(c), and can be considered as the principle difference of the SML and RSA regimes of carrier spin accumulation.  Note that the regime of weak dephasing is similar to the dynamics of a single spin presented in Figs.~\ref{fig:fig4} and \ref{fig:fig5}.

It is worth to mention, that the ratio of the signal amplitudes at negative and positive delays depends strongly on the generation efficiency and conservation of spin polarization, i.e., on the pump pulse area, the trion spin relaxation, and the ratio of carrier spin relaxation time $\tau_s$ to $T_R$~\cite{A.Greilich07212006,Gaby book}.

\section{RSA versus mode-locking}
\label{sec:rsa_vs_ml}

In this Section we discuss how one can distinguish the RSA and SML regimes and what parameters are responsible for separating these regimes. This separation is based on the common basic mechanism of the RSA and SML effects, which is the accumulation of carrier spin polarization under periodic pump pulse excitation. The key difference between the regimes is the ratio of the Larmor frequency broadening to the repetition frequency of the pump pulses: $\Delta \omega / \omega_R$. This is schematically illustrated in Fig.~\ref{fig:fig13}(a,b) by the frequency spectrum of the spin ensemble in a  finite magnetic field. Here few PSC modes satisfying Eq.~(\ref{PSC}) from
$(N-2)\omega_R$ to $(N+2)\omega_R$ are indicated in by the dashed vertical lines. 

In the RSA regime $\Delta \omega \ll \omega_R$ and only one PSC
mode (or even none) can fall into the distribution of Larmor
frequencies. When the PSC mode coincides with the distribution maximum, as it is shown in Fig.~\ref{fig:fig13}(a), one obtains a peak in the RSA spectrum. And when the overlap between the mode and the distribution is absent the RSA spectrum has minimum.

For the SML regime involvement of at least
two PSC modes is necessary. Therefore, the condition for this regime
is $\Delta \omega \gtrsim \omega_R$, see
Fig.~\ref{fig:fig13}(b). The calculations given in this Section show
that in fact the transition to the SML regime happens already for
$\Delta \omega \gtrsim 0.5\omega_R$, when the tails of the Larmor
frequency distribution overlap with more than one PSC mode.

\begin{figure}[hbt]
\includegraphics[width=1.\linewidth]{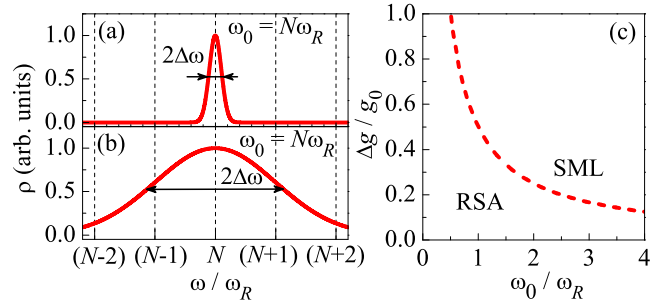}
\caption{Larmor frequency distribution function (multiplied for convenience
  by $\sqrt{2 \pi}\Delta \omega$) of a spin ensemble for RSA (a)
and SML (b) conditions. (c) Parameter diagram showing schematically
the regimes where RSA and SML occur, see text for details. } \label{fig:fig13}
\end{figure}

Deeper insight in the separation between the RSA and SML regimes is
collected below in Figs.~\ref{fig:fig14}, ~\ref{fig:fig15}, and
~\ref{fig:fig16}. Here the carrier spin polarization amplitude, $S_z^b$, and the signal phase at zero negative delay are analyzed as functions of magnetic
field, time delay, Larmor frequency spread, and pump pulse
area. We also consider the effect of resident carrier
  spin relaxation taking it into account via the parameter $\tau_s /
  T_R$. For most figures a pump pulse area $\Theta=\pi$ is
chosen as it provides efficient spin accumulation.
Let us go step by step through this data set.

First, for demonstration purposes, we assume again that a spread of the
carrier spin precession frequencies is $\Delta\omega=\omega_R$, and it
does not depend on magnetic field. For \textit{n}-type structures this corresponds to the case when the $\Delta\omega$ of the resident electrons is dominated by the random fields of the nuclear spin fluctuations: $\Delta \omega_{\rm n} \propto B_{\rm n,x}$. For $B>B_{\rm n}$ only the $B_{\rm n,x}$ component parallel to the external magnetic field should be considered, see Sec.~\ref{subsec:spread_nuclei}. Similar to the previous
Sections, the nuclear spin fluctuation is considered to be frozen.

Magnetic field dependencies of the carrier spin polarization, $-S_z^b$, and
the signal phase are shown in Figs.~\ref{fig:fig14}(a) and
\ref{fig:fig14}(b) for different $\Delta \omega$ and
$\tau_s/T_R=300$. For a small frequency spread of $\Delta \omega=0$ and
$0.2\omega_R$ the polarization amplitude and phase are periodic
functions of magnetic field, which is characteristic for the RSA
regime, for comparison see Figs.~\ref{fig:fig5}(b) and
\ref{fig:fig5}(d). An increase of $\Delta \omega$ to $0.5\omega_R$ drastically changes the character of these functions:
both of them become independent of magnetic field. The spin
polarization amplitude has a finite value (in this case it is equal
0.08), while $\phi=0$. These are characteristics of the SML regime.

\begin{figure}[hbt]
\includegraphics[width=1.\linewidth]{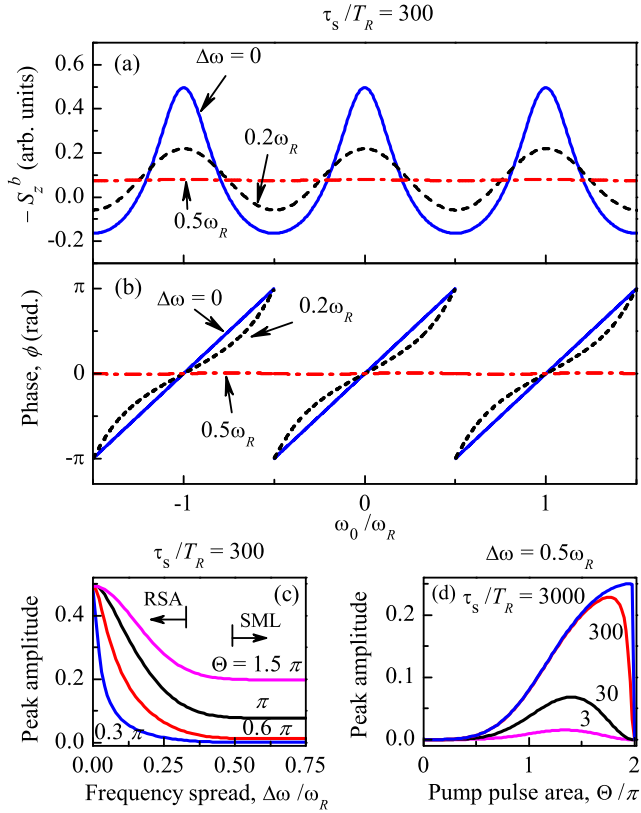}
\caption{(Color online) Magnetic field dependencies [in terms of
  $\omega_0(B) / \omega_R$] of (a) the carrier spin polarization amplitude,
  $-S_z^b$, and (b) the signal phase at zero delay calculated for three
  different Larmor frequency spreads. Dependencies of the spin
  polarization amplitude for PSC modes, i.e. for integer values of
  $\omega_0(B) / \omega_R$ (c) on the frequency spread for different
  pump pulse areas, at $\tau_s/T_R=300$; and  (d) on the pump pulse
  area for various $\tau_s/T_R$, for a precession frequency spread $\Delta \omega = 0.5
  \omega_R$.} \label{fig:fig14}
\end{figure}

Details of separating the RSA from the SML regime with increasing
frequency spread are presented in Fig.~\ref{fig:fig14}(c). The peak
amplitudes of the spin polarization at the PSC frequencies are plotted for different pump pulse areas there. The amplitude initially decreases with an increasing spread and approaches a saturation level for larger
spreads. Independence of the amplitude on the spread is characteristic
for the SML regime, therefore, one can see from the Fig.~\ref{fig:fig14}(c) that the
regimes cross over at $\Delta \omega \sim  0.5 \omega_R$. 

The
spin polarization amplitude in the SML regime depends critically on the
pump pulse area, see also Fig.~\ref{fig:fig14}(d). It is close to
zero for $\Theta < 0.3\pi$, but strongly increases for $\Theta$
exceeding this value, approaching a maximum at $\Theta = 2\pi$ for
sufficiently large $\tau_s/T_R = 3000$. The dependence of $S_z^b$ for a large spread, which
corresponds to a constant plateau level, can be written as:
\begin{equation}
 S_z^b  =\frac{1-Q}{1+Q}\left[1-\sqrt{\frac{M^2-1}{L^2-1}} \right],
\end{equation}
where $M=Q\mathrm e^{-T_R/\tau_s}$ and $L=\mathrm
  e^{-T_R/\tau_s}(1+Q^2)/2$. The calculations in Fig.~\ref{fig:fig14}(d) show that with increasing electron spin relaxation time $\tau_s$ the maximum signal amplitude shifts to a pulse area of 2$\pi$ [unlike the dependence of spin polarization on pulse area for excitation by a single pulse, for which Rabi oscillations occur with maximum at $\Theta=\pi$].

The fact that the separation between RSA and SML is controlled by
the ratio $\Delta \omega / \omega_R$ offers the instructive
opportunity to realise a changeover between these two regimes by
tuning the magnetic field. This would be possible for the case when
the Larmor frequency spread is controlled by $\Delta g$, see
Sec.~\ref{subsec:spread}, because in this case $\Delta \omega_g$
increases linearly with $B$. Results of corresponding calculations
for $\Delta \omega_g = 0.1 \omega_0$  are given in
Fig.~\ref{fig:fig15}. In analogy with Figs.~\ref{fig:fig14}(a) and
\ref{fig:fig14}(b), one can identify the RSA regime in low magnetic
fields ($|\omega_0 / \omega_R | < 3$), where both the polarization
amplitude and the phase change with $B$, and the SML regime in
larger magnetic fields ($|\omega_0 / \omega_R | > 5$), where these
parameters do not vary anymore.

Figure~\ref{fig:fig13}(c) shows the range of parameters in which the
different spin accumulation regimes can be obtained. The dashed
curve corresponds to the condition $\Delta \omega=0.5\omega_R$, which
may serve as approximate boundary between the RSA and SML regimes.
Indeed, if the $g$ factor spread is small, the spin frequency distribution contains only one phase
synchronized mode in a broad range of magnetic fields, the latter are expressed via $\omega_0(B)/\omega_R$. It corresponds to the RSA regime for which the parameter space is placed below the dashed curve in
Fig.~\ref{fig:fig13}(c). On the contrary, if the $g$-factor spread
is large, several phase synchronized modes become involved already at weak magnetic fields and, for
relatively efficient optical pumping, SML occurs [the parameter
space above the dashed curve].

\begin{figure}[hbt]
\includegraphics[width=1.\linewidth]{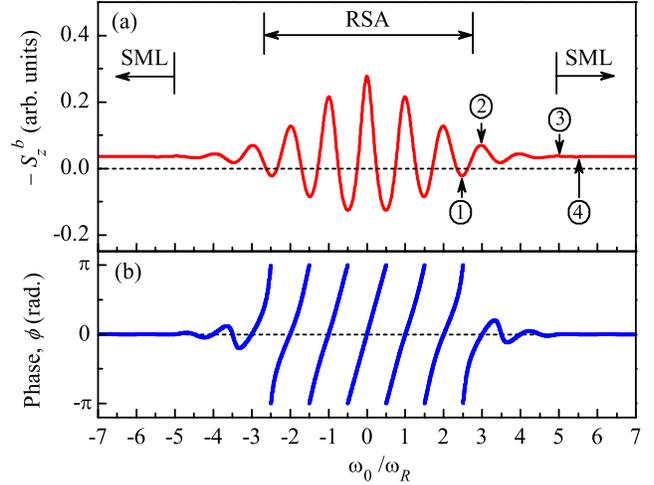}
\caption{(Color online) Magnetic field dependence of (a) carrier
spin
  polarization $-S_z^b$ at zero pump-probe delay (shortly before pump
  pulse arrival), and (b) spin precession phase of the
  signal calculated for the same parameters as in panel (a). The RSA and
  SML regimes are shown by arrows. The labels with numbers are in accordance with Fig.~\ref{fig:fig16}. $\tau_s/T_R=3$, $\Theta =
  \pi$, $\Delta \omega_g = 0.1 \omega_0$.}
\label{fig:fig15}
\end{figure}

The time evolution of the spin polarization for the magnetic fields
in Fig.~\ref{fig:fig15}(a) are given in Fig.~\ref{fig:fig16}. Panel
(a) corresponds to the RSA regime (weak magnetic fields). One can see
that the spin polarization phase and amplitude at small negative delays
depends on the relation to the PSC. However, in the ML regime, shown
in panel (b), both values are constant irrespective whether the PSC
are fulfilled or not.

\begin{figure}[hbt]
\includegraphics[width=1.\linewidth]{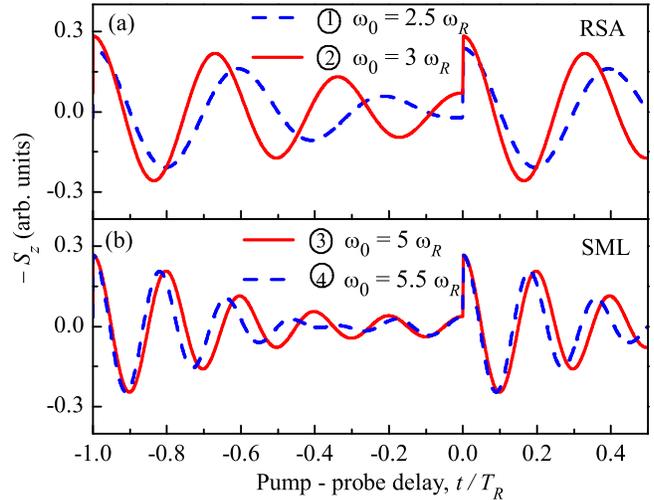}
\caption{(Color online) Carrier spin polarization as function of
  pump-probe delay for different magnetic fields denoted
  in Fig.~\ref{fig:fig15}(a). Panel (a) corresponds to the RSA regime,
  and panel (b) to the SML regime. $\tau_s/T_R=3$, $\Theta = \pi$,
  $\Delta \omega_g = 0.1 \omega_0$.} \label{fig:fig16}
\end{figure}

From the results of Secs.~\ref{sec:modelock} and \ref{sec:rsa_vs_ml}
one can conclude about the two main features of the SML regime. The
first one is a fixed phase of the spin signal at very small negative delays, which is independent of the
magnetic field. This reflects the
primary amplification of spins with commensurable spin beat
frequencies in a strongly inhomogeneous ensemble. The second one is
a characteristic revival of the dephased signal before the next pump
pulse arrival shown in Fig.~\ref{fig:fig12}.

It is also interesting, that contrary to the RSA regime in the SML
regime the magnetic field dependence of the spin polarization at
zero negative delay is smooth. The dependence is similar to that
presented by the dashed line in Fig.~\ref{fig:fig11}(c). The width
of this bell-like curve is determined by the nuclear field
fluctuations and is approximately equal to $4 \Delta_B$.

Let us summarize the conditions for the SML regime. Apart from the
obvious condition $\tau_s \gg T_R$ it requires:
\begin{enumerate}
\item A significant spread of carrier spin precession frequencies, $\Delta \omega_g >
  0.5 \omega_R$. The spread can be caused by the nuclear
  fluctuation fields or by the spread of $g$ factors.
\item The frequency spread $\Delta \omega > 0.5 \omega_R$ leads to a
  dephasing of the spin signal within the time $T_2^* \sim T_R/\pi$,
  {i.e. faster than the time interval between subsequent pump
    pulses}.
\item One can see from Figs.~\ref{fig:fig14}(c) and \ref{fig:fig14}(d)
  that the pump pulse area should be sufficiently large, $\Theta \gtrsim
  \pi/2$. Otherwise the frequency spread $\Delta \omega > 0.5 \omega_R$ would
  cause only a decay of the spin polarization without its revival before
  the next pump pulse arrival.
\end{enumerate}

 \section{Conclusions}\label{sec:concl}

To conclude, we have performed a comprehensive theoretical study  of
carrier spin coherence in spin ensembles subject to periodic optical
pumping.  The effect of spin accumulation has been analysed for
singly-charged quantum dots and quantum wells with a low density
carrier gas. The accumulation results in two regimes of carrier spin
coherence: resonance spin amplification and spin mode-locking. These
regimes, while being different in their phenomenological appearances
and realization conditions, have the same origin and occur for spin
ensembles for which the carrier spin coherence time exceeds by far
the pump repetition period. The resonance spin amplification and
spin mode-locking are mutually exclusive regimes because of the
requirement on excitation power and precession frequency spread.

For the RSA regime sufficiently homogeneous spin ensembles and small
excitation powers (small pump pulse areas) are required. These
conditions are experimentally realized in QW structures with
electron or hole resident carriers of low density, i.e. for the
regime, where negatively or positively charged trions play an
important role. In this case the spin dephasing times for resident
carriers can be extracted with high accuracy, even when they exceed
the pulse repetition period. The spreads of $g$ factors and nuclear
spin fluctuations are less important for the long-lived spin
coherence compared to the case of strongly inhomogeneous QD
ensembles.

In contrast to the RSA regime the SML regime requires a strong
inhomogeneity of the spin precession frequency in the spin ensemble
and high excitation powers (pump areas close to $\pi$ and more). By now the
SML regime has been observed experimentally and studied in great
detail for ensembles of (In,Ga)As/GaAs QDs each singly charged with
a resident electron. In principle it may be also observable for quantum dots
singly-charged with a resident hole, if the respective
conditions are met.

\acknowledgments

The authors thank A. Greilich, Al. L. Efros, I. V. Ignatiev,
and E. L. Ivchenko for valuable discussions. This work was supported
by the Deutsche Forschungsgemeinschaft, the Russian Foundation of Basic
Research and the EU Seventh
Framework Programme (Grant No. 237252, Spin-optronics). IAY is a Fellow of the Alexander von Humboldt Foundation. MMG acknowledges support of the
``Dynasty'' Foundation---ICFPM.


\begin{thebibliography}{34}%

%
\bibitem{Awschalom_Spintronics} {{\em Semiconductor Spintronics and Quantum Computation}, edited by D. D. Awschalom, D. Loss, and N. Samarth (Springer, Berlin, 2002).}
%
\bibitem{Gabi_book_1} {{\em Optical Generation and Control of Quantum Coherence in Semiconductor Nanostructures}, Springer Series in NanoScience and Technology, edited by G. Slavcheva and Ph. Roussignol (Springer, Berlin 2010). ISBN: 978-3-642-12490-7.}

%
\bibitem{Fritz_book} {{\em Semiconductor Quantum Bits}, edited by F. Henneberger and O. Benson (Pan Stanford Publishing, Singapore, 2009). ISBN: 978-981-4241-05-2}%
%
\bibitem{Dyakonov_Spin} {\em Spin Physics in Semiconductors}, edited by M. Dyakonov (Springer, Berlin 2008). ISBN: 978-3-540-78819-5.
%
\bibitem{Kikkawa98} J. M. Kikkawa and D. D. Awschalom, Phys. Rev. Lett. {\bf 80}, 4313 (1998).
%
\bibitem{Kikkawa_Science97} J. M. Kikkawa, I. P. Smorchkova, N. Samarth, and D. D. Awschalom, {Science} {\bf 277}, 1284 (1997).
%
\bibitem{GaN} B. Beschoten, E. Johnston-Halperin, D. K. Young, M. Poggio, J. E. Grimaldi, S. Keller, S. P. DenBaars, U. K. Mishra, E. L. Hu, and D. D. Awschalom,
Phys. Rev. B {\bf 63}, 121202 (2001).
%
\bibitem{kennedy:045307} T. A. Kennedy, A. Shabaev, M. Scheibner, A. L. Efros, A. S. Bracker, and D. Gammon, Phys. Rev. B {\bf 73}, 045307 (2006).
%
\bibitem{zhu07} E. A. Zhukov, D. R. Yakovlev, M. Bayer, M. M. Glazov, E. L. Ivchenko, G. Karczewski, T. Wojtowicz, and J. Kossut, Phys. Rev. B {\bf 76}, 205310 (2007).
%
\bibitem{QW} W. J. H. Leyland, G. H. John, R. T. Harley, M. M. Glazov, E. L. Ivchenko, D. A. Ritchie, I. Farrer, A. J. Shields, and M. Henini, Phys. Rev. B \textbf{75}, 165309 (2007).
%
\bibitem{Gupta_PRB99} J. A. Gupta, D. D. Awschalom, X. Peng, and A. P. Alivisatos, Phys. Rev. B
{\bf 59} R10421 (1999). 
\bibitem{Gupta_PRB02} J. A. Gupta, D. D. Awschalom, Al. L. Efros, and A. V. Rodina, 
Phys. Rev. B {\bf 66}, 125307 (2002).
%
\bibitem{Petroff_APL01} R. Epstein, D. T. Fuchs, W. V. Schoenfeld, P. M. Petroff, and D. D.
Awschalom, Appl. Phys. Lett. {\bf 76}, 733 (2001).
%
\bibitem{A.Greilich07212006} A. Greilich, D. R. Yakovlev, A. Shabaev, A. L. Efros, I. A. Yugova, R. Oulton, V. Stavarache, D. Reuter, A. Wieck, and M. Bayer, Science {\bf 313}, 341 (2006).
%
\bibitem{greilich06} A. Greilich, R. Oulton, E. A. Zhukov, I. A. Yugova, D. R. Yakovlev, M. Bayer, A. Shabaev, A. L. Efros, I. A. Merkulov, V. Stavarache, D. Reuter, and A. Wieck, Phys. Rev. Lett. {\bf 96}, 227401 (2006).
%
\bibitem{Greilich_PRB07}A. Greilich, M. Wiemann, F. G. G. Hernandez, D. R. Yakovlev, I. A. Yugova, M. Bayer, A. Shabaev, Al. L. Efros, D. Reuter, and A. D. Wieck, Phys. Rev. B. {\bf 75}, 233301 (2007).
%
\bibitem{OptOr} {{\em Optical Orientation}, eds. F. Meier and
B. P. Zakharchenya (North-Holland, Amsterdam, 1984).}
%
\bibitem{Ber08} {J. Berezovsky, M. H. Mikkelsen, N. G. Stoltz, L. A. Coldren, and D. D. Awschalom, Science \textbf{320}, 349 (2008).}
%
\bibitem{Atature07} M. Atature, J. Dreiser, A. Badolato, and A. Imamoglu,
Nature Phys. \textbf{3}, 101 (2007).
%
\bibitem{Gaby book} A. Greilich, D. R. Yakovlev and M. Bayer, Chapter 6 on {\em Ensemble spin coherence of singly charged InAs quantum dots}, pp. 85-127 in book {\em Optical Generation and Control of Quantum Coherence in Semiconductor Nanostructures}, ed. by G. Slavcheva and Ph. Roussignol (Springer, Berlin 2010). ISBN: 978-3-642-12490-7.
%
\bibitem{chapter6} D. R. Yakovlev and M. Bayer, Chapter 6 on {\em Coherent spin dynamics of carriers}, pp.135-177 in book {\em Spin Physics in Semiconductors}, ed. by M. I. Dyakonov (Springer, Berlin 2008). ISBN: 978-3-540-78819-5.
%
\bibitem{fokina-2010} L. V. Fokina, I. A. Yugova, D. R. Yakovlev, M. M. Glazov, I. A. Akimov, A. Greilich, D. Reuter, A. D. Wieck, and M. Bayer, Phys. Rev. B {\bf 81}, 195304 (2010).
%
\bibitem{yugova09} I. A. Yugova, M. M. Glazov, E. L. Ivchenko, and A. L. Efros, Phys. Rev. B {\bf 80}, 104436 (2009).
%
\bibitem{longpulse11} S. Spatzek, S. Varwig, M. M. Glazov,
I. A. Yugova, A. Schwan, D. R. Yakovlev, D. Reuter, A. D. Wieck,
and M. Bayer, Phys. Rev. B {\bf 84}, 115309 (2011).

\bibitem{reviewFTT} M. M. Glazov, Fiz. Tverd. Tela {\bf 54},
3 (2012) [Engl. Transl.: Phys. Solid State (2012), in press].
%
\bibitem{shabaev:201305} A. Shabaev, A. L. Efros, D. Gammon, and I. A. Merkulov, Phys. Rev. B {\bf 68}, 201305 (2003).
%
\bibitem{bonadeo98} N. H.~Bonadeo, J.~Erland, D.~Gammon, D.~Park, D. S.~Katzer and D. G.~Steel, Science {\bf 282}, 1473 (1998).
%
\bibitem{zhukov10} E. A. Zhukov, D. R. Yakovlev, M. M. Glazov, L. Fokina, G. Karczewski, T. Wojtowicz, J. Kossut, and M. Bayer, Phys. Rev. B {\bf 81}, 235320 (2010).
%
\bibitem{comment} An ensemble of electron spins in magnetic field
can be characterized by three relaxation times [see A. Abragam, {\em
The Principles of Nuclear Magnetism} (Oxford University Press,
Oxford, 1961), p. 44.]: the longitudinal spin relaxation time $T_1$,
which is related to the relaxation of the spin component parallel to
the field ($S_x$ in our case), and the transverse spin relaxation times
$T_2$ and $T_2^*$ . The $T_2$ time describes spin decoherence (i.e.,
relaxation of the spin components transverse to the field: $S_y$ and
$S_z$ in our notation) of a single carrier, while the $T_2^*$ time
describes the dephasing of the spin ensemble (e.g., due to the
inhomogeneous broadening of the carrier $g$ factor). Due to the
anisotropy of spin relaxation in QWs
\cite{dyakonov86,averkiev:15582,averkiev06,larionov:033302} the spin
relaxation times $T_y$ and $T_z$ differ from each other. In
Eqs.~(\ref{system}) we assume for simplicity an isotropic relaxation
time: $\tau_s=T_x=T_y=T_z$. The effect of spin dephasing anisotropy
is considered in Sec.~\ref{subsec:rsa_any}.
%
\bibitem{PhysRevB.75.115330} I. Y. Gerlovin, Y. P. Efimov, Y. K. Dolgikh, S. A. Eliseev, V. V. Ovsyankin, V. V. Petrov, R. V. Cherbunin, I. V. Ignatiev, I. A. Yugova, L. V. Fokina, A. Greilich, D.R. Yakovlev, and M. Bayer, Phys. Rev. B {\bf 75}, 115330 (2007).
%
\bibitem{com5} The situation is typical for \textit{n}-type structures with resident electrons, where the T$^-$ trion spin dynamics are controlled by the hole spin.
This is related to the fact that in QWs and epitaxially grown QDs
the in-plain \textit{g} factor of the heavy-hole is close to zero
and, therefore, one can neglect the hole Larmor precession, i.e.
$\Omega=0$, to a good approximation.
%
\bibitem{marie99} X. Marie, T. Amand, P. Le Jeune, M. Paillard, P. Renucci, L. E. Golub, V. D. Dymnikov, and E. L. Ivchenko, Phys. Rev. {\bf 60}, 5811 (1999).
%
\bibitem{yugova02} I. A. Yugova, I. Ya. Gerlovin, V. G. Davydov, I. V. Ignatiev, I. E. Kozin, H.-W. Ren, M. Sugisaki, S. Sugou, and Y. Masumoto, Phys. Rev. B {\bf 66}, 235312 (2002).
%
\bibitem{stevenson} R. M. Stevenson, R. J. Young, P. See, D. G. Gevaux, K. Cooper, P. Atkinson, I. Farrer, D. A. Ritchie, and A. J. Shields, Phys. Rev. B {\bf 73}, 033306  (2006).
%
\bibitem{Machnikowski10} P. Machnikowski and T. Kuhn, Phys. Rev. B {\bf 81}, 115306 (2010).
%
\bibitem{korn_njp} T. Korn, M. Kugler, M. Griesbeck, R. Schulz, A. Wagner, M. Hirmer, C. Gerl, D. Schuh, W. Wegscheider, and C. Sch{\"u}ller, New Journal of Physics {\bf 12}, 043003 (2010).
%
\bibitem{beschoten} B. Beschoten, {\em Spin Coherence in Semiconductors} in {\em Magnetism goes Nano}, 36th Spring School 2005, Schriften des Forschungzentrums J\"ulich, Matter and Materials, vol. 26  (2005), edited by T. B. S. Blugel and C. Schneider, pp. E7.1 - E7.27.
%
\bibitem{glazov08a} M. M. Glazov and E. L. Ivchenko, Sov. Phys. Semicond. {\bf 42}, 951 (2008).
%
\bibitem{comment_t2}
Strictly speaking, the dephasing time of a spin ensemble, $T_2^*$,
is magnetic field dependent due to the spread of carrier $g$
factors, see Sec.~\ref{subsec:dephasing}.  The effect of the peak
broadening for an increasing  magnetic field (that is with an
increase of the peak number $N$) is analyzed in
Sec.~\ref{subsec:dephasing}.
%
\bibitem{ast08} G. V. Astakhov, M. M. Glazov, D. R. Yakovlev, E. A. Zhukov, W. Ossau, L. W. Molenkamp, and M. Bayer, Semicond. Science and Technology {\bf 23}, 114001 (2008).
%
\bibitem{zhu06} E. A. Zhukov, D. R. Yakovlev, M. Bayer, G. Karczewski, T. Wojtowicz, and J. Kossut, Phys. Stat. Sol. B {\bf 243}, 878 (2006).
%
\bibitem{sokolova09} I. A. Yugova, A. A. Sokolova, D. R. Yakovlev, A. Greilich, D. Reuter, A. D. Wieck, and M. Bayer, Phys. Rev. Lett. {\bf 102}, 167402 (2009).
%
\bibitem{greilich09} A. Greilich, S. Spatzek, I. A. Yugova, I. A. Akimov, D. R. Yakovlev, A. L. Efros, D. Reuter, A. D. Wieck, and M. Bayer, Phys. Rev. B. {\bf 79}, 201305(R) (2009).
%
\bibitem{Kugler11} M. Kugler, K. Korzekwa, P. Machnikowski, C. Gradl, S. Furthmeier, M. Griesbeck, M. Hirmer, D. Schuh, W. Wegscheider, T. Kuhn, C. Sch\"{u}ller, and T. Korn, Phys. Rev. B {\bf 84}, 085327 (2011).
%
\bibitem{dyakonov86} M. Dyakonov and V. Kachorovskii, Sov. Phys. Semicond. {\bf 20}, 110 (1986).
%
\bibitem{averkiev:15582} N. S. Averkiev and L. E. Golub, Phys. Rev. B {\bf 60}, 15582 (1999).
%
\bibitem{averkiev06} N. S. Averkiev, L. E. Golub, A. S. Gurevich, V. P. Evtikhiev, V. P. Kochereshko, A. V. Platonov, A. S. Shkolnik, and Y. P. Efimov, Phys. Rev. B {\bf 74}, 033305 (2006).
%
\bibitem{larionov:033302} A. V. Larionov and L. E. Golub, Phys. Rev. B {\bf 78}, 033302 (2008).
%
\bibitem{willander} N. S. Averkiev, L. E. Golub, and
    M. Willander, J. Phys.: Condens. Matter  {\bf 14}, R271 (2002).
%
\bibitem{Griesbeck11} M. Griesbeck, M. M. Glazov,
    E. Ya. Sherman, D. Schuh, W. Wegscheider, C. Sch\"uller, T. Korn,
    preprint arXiv:1111.5438 (2011).
%
\bibitem{Ignatiev05} M. Ikezawa, B. Pal, Y. Masumoto, I. V. Ignatiev, S. Y. Verbin, and I. Y. Gerlovin, Phys. Rev. B {\bf 72}, 153302 (2005).
%
\bibitem{PhysRevB.64.125316} A. V. Khaetskii and Y. V. Nazarov, Phys. Rev. B {\bf 64}, 125316 (2001).
%
\bibitem{PhysRevB.66.161318} L. M. Woods, T. L. Reinecke, and Y. Lyanda-Geller, Phys. Rev. B {\bf 66}, 161318 (2002).
%
\bibitem{carter:167403} S. G. Carter, A. Shabaev, S. E. Economou, T. A. Kennedy, A. S. Bracker, and T. L. Reinecke, Phys. Rev. Lett. {\bf 102}, 167403 (2009).
%
\bibitem{ivchenko05a} E. L. Ivchenko, {\it Optical Spectroscopy of Semiconductor Nanostructures} (Alpha Science, Harrow UK, 2005).
%
\bibitem{PhysRevB.75.245302} I. A. Yugova, A. Greilich, D. R. Yakovlev, A. A. Kiselev, M. Bayer, V. V. Petrov, Y. K. Dolgikh, D. Reuter, and A. D. Wieck, Phys. Rev. B {\bf 75}, 245302 (2007).
%
\bibitem{Kotlyar01} R. Kotlyar, T. L. Reinecke, M. Bayer, and A. Forchel, Phys. Rev. B {\bf 63}, 085310 (2001).
%
\bibitem{merkulov02} I. A. Merkulov, A. L. Efros, and M. Rosen, Phys. Rev. B {\bf 65}, 205309 (2002).
%
\bibitem{syperek11} M. Syperek, D. R. Yakovlev, I. A. Yugova, J. Misiewicz, I. V. Sedova, S. V. Sorokin, A. A. Toropov, S. V. Ivanov, and M. Bayer, Phys. Rev. B {\bf 84}, 085304 (2011) and Phys. Rev. B {\bf 84}, 15990(E) (2011).
%
\bibitem{comment_dephasing} If the spreads of $g$ factors and nuclear
  spin fluctuation fields are described by normal distributions
  (Gaussians) then the dispersion of electron-spin precession
  frequencies in the QD ensemble is \cite{greilich09}: $\Delta
  \omega=\sqrt{(\Delta g\mu_BB/\hbar)^2+\omega^2_{\rm
      n}}$. This leads to a Gaussian dephasing with a
  characteristic time $T^*_{2}=1/\Delta \omega$.
%
\bibitem{petrov08} M. Yu. Petrov, I. V. Ignatiev, S. V. Poltavtsev, A. Greilich,
A. Bauschulte, D. R. Yakovlev, and M. Bayer, Phys. Rev. B {\bf 78}, 045315 (2008).
%
\bibitem{comment_hanle}
The zeroth RSA peak [Eq.~\eqref{Lorentian0}] has the same form as
the standard expression for the Hanle effect \cite{OptOr}: the
electron spin depolarization in a transversal magnetic field under
continuous wave pumping. The influence of the inhomogeneous
distribution of $g$ factors on the Hanle effect is typically quite
weak and, as a rule~\cite{ast08,sandreev}, the extracted spin
dephasing time is controlled by the nuclear spin fluctuations, i.e.,
by $T_2^*$. As compared with the Hanle effect, the studies of the
resonant spin amplification allow one to directly extract the
magnetic field dependence of the spin dephasing time, and
consequently, evaluate the spread of $g$ factors, $\Delta g$.
%
\bibitem{sandreev} {S. V. Andreev, B. R. Namozov,
A. V. Koudinov, Yu. G. Kusrayev, and J. K. Furdyna, Phys. Rev. B
{\bf 80}, 113301 (2009).}
%

\end{thebibliography}
\end{document}